\documentclass[aps,prx,reprint,superscriptaddress,longbibliography]{revtex4-2}

\usepackage{graphicx}
\usepackage{color,amssymb,amsmath}
\usepackage[normalem]{ulem}
\usepackage{hyperref}
\usepackage{lipsum}
\usepackage{physics}
\usepackage{float}

\makeatletter
\newcommand*{\rom}[1]{\expandafter\@slowromancap\romannumeral #1@}
\makeatother

\begin{document}
\title{Finite Length Effects and Coulomb Interaction in Ge Quantum Well-Based Josephson Junctions Probed with Microwave Spectroscopy}

\author{S.~C.~ten~Kate}
\affiliation{IBM Research Europe - Zurich, S\"aumerstrasse 4, 8803 R\"uschlikon, Switzerland}

\author{D.~C.~Ohnmacht}
\affiliation{Fachbereich Physik, Universit\"at Konstanz, D-78457, Konstanz, Germany}
	
\author{M.~Coraiola}
\affiliation{IBM Research Europe - Zurich, S\"aumerstrasse 4, 8803 R\"uschlikon, Switzerland}

\author{T.~Antonelli}
\affiliation{IBM Research Europe - Zurich, S\"aumerstrasse 4, 8803 R\"uschlikon, Switzerland}

\author{S.~Paredes}
\affiliation{IBM Research Europe - Zurich, S\"aumerstrasse 4, 8803 R\"uschlikon, Switzerland}

\author{F.~J.~Schupp}
\affiliation{IBM Research Europe - Zurich, S\"aumerstrasse 4, 8803 R\"uschlikon, Switzerland}

\author{M.~Hinderling}
\affiliation{IBM Research Europe - Zurich, S\"aumerstrasse 4, 8803 R\"uschlikon, Switzerland}

\author{S.~W.~Bedell}
\affiliation{IBM Quantum, T.J. Watson Research Center, 1101 Kitchawan Road, Yorktown Heights, New York 10598, USA}

\author{W.~Belzig}
\affiliation{Fachbereich Physik, Universit\"at Konstanz, D-78457, Konstanz, Germany}

\author{J.~C.~Cuevas}
\affiliation{Departamento de F\'isica Te\'orica de la Materia Condensada and Condensed Matter Physics Center (IFIMAC), Universidad Aut\'onoma de Madrid, E-28049, Madrid, Spain}

\author{A.~E.~Svetogorov}
\affiliation{Fachbereich Physik, Universit\"at Konstanz, D-78457, Konstanz, Germany}

\author{F.~Nichele}
\email{fni@zurich.ibm.com}
\affiliation{IBM Research Europe - Zurich, S\"aumerstrasse 4, 8803 R\"uschlikon, Switzerland}

\author{D.~Sabonis}
\email{Deividas.Sabonis@ibm.com}
\affiliation{IBM Research Europe - Zurich, S\"aumerstrasse 4, 8803 R\"uschlikon, Switzerland}

\date{\today}

\begin{abstract}
Proximitized Ge quantum wells have emerged as a novel platform for studying Andreev bound states (ABSs), due to their expected strong spin-orbit interaction and high mobility. Here, we used microwave spectroscopy techniques to investigate ABSs in Josephson junctions (JJs) realized in proximitized Ge quantum wells. Spectroscopic signatures observed in a 350 nm junction indicated the presence of multiple ABSs, and were reproduced with a model including finite-length effects. The ABS spectra measured for a $1.2~\mu$m junction were explained by a model including three ABSs in two conduction channels and finite Coulomb interaction. Our work highlights the importance of interactions in JJs and serves as a basis for understanding and manipulating ABSs in Ge-based hybrid devices.
\end{abstract}

\maketitle

\section*{Introduction}
A semiconductor confined between two superconductors forms a Josephson junction (JJ), where discrete sub-gap states, known as Andreev bound states (ABSs) arise, which are responsible for the current transport across the JJ~\cite{Andreev1964,BTK1982,Beenakker1991_2,Furusaki1991}. These states have successfully been probed using spectroscopy techniques in dc transport~\cite{Pillet2013,Nichele2020,Coraiola2023} and microwave~\cite{Janvier2015,Hays2018,Tosi2019,Hays2020,Hays2021,Chidambaram2022,Canadas2022,Zellekens2022,Fatemi2022,Hinderling2023,Wesdorp2024,Hinderling2024Ge,Elfeky2025} domain experiments, where effects of Coulomb~\cite{Kurilovich2021,Canadas2022,Fatemi2022,Kurilovich2024}, exchange~\cite{Wesdorp2024} and spin-orbit~\cite{Tosi2019,Hays2020,Hays2021,Canadas2022,Zellekens2022,Wesdorp2024,Elfeky2025} interaction have been demonstrated for InAs nanowire-based junctions. Recently, a hard induced superconducting gap has been demonstrated in Ge quantum wells~\cite{Tosato2023}, forming a novel platform for studying ABSs. It has been shown that the two-dimension hole gases (2DHGs) formed in these quantum wells have high mobility~\cite{Lodari2019} and strong spin-orbit interaction~\cite{Scappucci2021}. Moreover, isotopic purification provides the opportunity to create a nuclear-spin-free environment in this material platform, thus mitigating existing limitations towards spin-based quantum information processing applications. For these reasons, Ge quantum wells constitute a promising platform for realizing semiconductor spin qubits and hybrid Andreev qubits~\cite{Scappucci2021}. Initial investigations of ABSs in Ge quantum wells have revealed the feasibility of this platform for realizing hybrid quantum devices~\cite{Hinderling2024Ge,Lakic2025}. However, the physical mechanisms affecting ABS energies in Ge-based JJs are not yet well understood.

Here, we utilized microwave spectroscopy techniques to study ABSs in JJs of various length realized in Ge quantum wells. The junctions were embedded in superconducting loops, which were inductively coupled to Nb coplanar waveguide resonators using the flip-chip approach~\cite{Zellekens2022,Hinderling2023,Hinderling2024,Hinderling2024Ge}. This technique enabled us to reduce dielectric losses by separating the superconducting resonators from the Ge heterostructure with a vacuum gap, resulting in high resonator internal quality factors. Improved readout sensitivity compared to our previous work~\cite{Hinderling2024Ge} enabled observation of novel features in microwave spectroscopy measurements, namely single-quasiparticle transitions (SQPTs) and mixed pair transitions (mixed PTs). The experimentally measured ABS spectra were reproduced by models incorporating multiple interacting ABSs in the junction, providing good qualitative agreement. For a junction of length $L$ shorter than the superconducting coherence length of Ge $\xi_\mathrm{Ge}$, ABS spectra contained multiple pair transitions (PTs) and an SQPT. Moreover, in a long junction with $L>\xi_\mathrm{Ge}$, many PTs and SQPTs were observed. Coulomb interaction shifted the transition frequencies depending on the quasiparticle (QP) occupation of ABSs, lifting the degeneracy between two SQPTs. These results highlight the importance of finite-length effects and Coulomb interaction in JJs realized in Ge quantum wells. 

\begin{figure*}
	\includegraphics[width=0.95\textwidth]{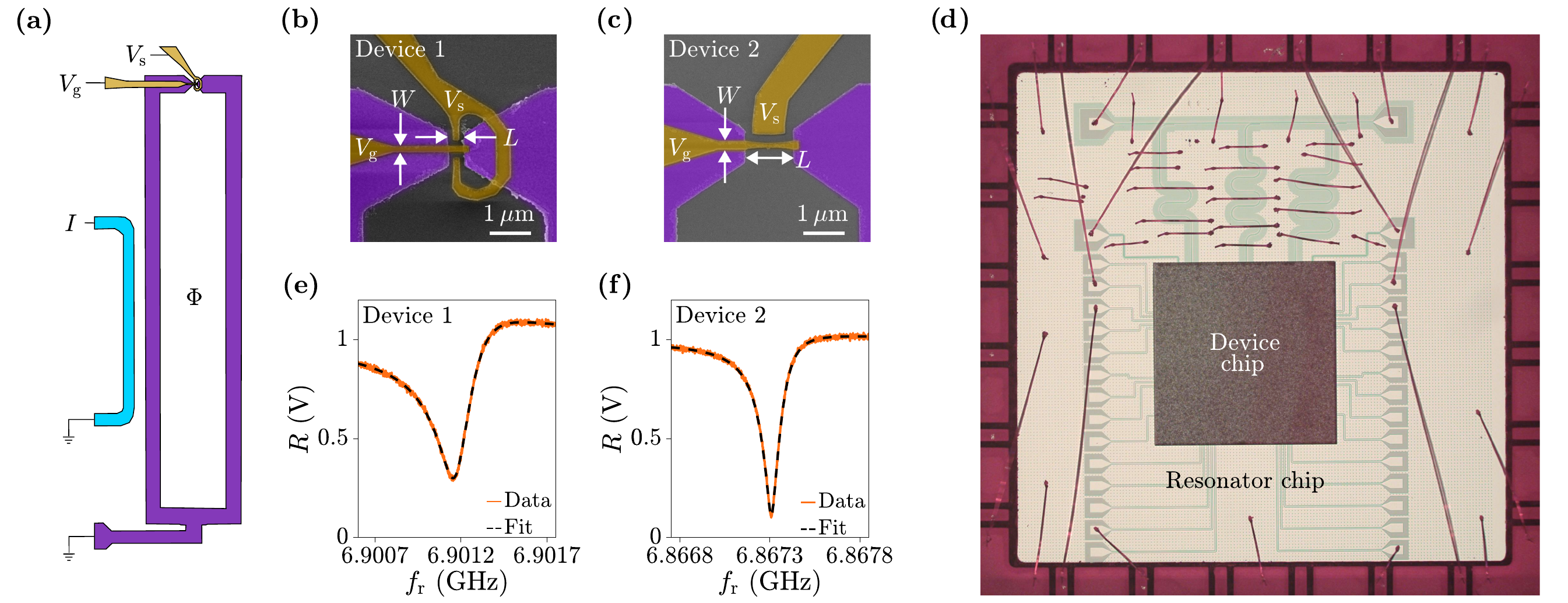}
	\caption{(a) Schematic of device~1, consisting of a PtSiGe superconducting loop (purple), Ti/Al/Ti/Au gates (yellow) and a Ti/Al/Ti/Au flux line (blue). The dc voltage $V_\mathrm{g}$ controlled the hole density in the Ge quantum well via the accumulation gate. The rf drive tone for two-tone spectroscopy measurements was applied via the side gate indicated with $V_\mathrm{s}$ using a bias-tee. For device 1, the magnetic flux $\Phi$ was controlled by passing a current $I$ through the flux line, while for device 2, $\Phi$ was controlled via a superconducting coil, mounted in the vicinity of the sample. (b,c) False-colored scanning electron micrographs (top view) of exemplary Josephson junctions, identical to device~1 (b) and 2 (c). The dimensions of each junction are defined by the separation $L$ between the superconducting PtSiGe electrodes (purple) and the width $W$ of the metallic accumulation gate (yellow). The side gate (yellow) was used to apply the rf drive tone. The gates were separated from the Ge heterostructure by a SiO${}_\mathrm{x}$ dielectric layer (not visible). (d) Optical microscope image after flip-chip bonding and wire bonding. The device chip ($3\times 3~\mathrm{mm}^2$) hosting the Ge quantum well was connected to the resonator chip ($10\times 10~\mathrm{mm}^2$) via In bumps. The grounded ends of the $\lambda/4$ Nb resonators were aligned to the superconducting loops of the devices (not visible). (e,f) Amplitude $R=\sqrt{I^2+Q^2}$ of the resonator transmission $S_{21}$ (normalized) as a function of readout frequency $f_\mathrm{r}$ for device 1~(e) and device~2~(f). The dashed black lines are fits to the data obtained using the circle-fit method~\cite{Probst2015}.}
	\label{fig1}
\end{figure*}

\section*{Devices and resonator characterization}

Two devices, named device~1 and device~2, were realized in strained Ge quantum wells grown with reduced-pressure chemical-vapor-deposition~\cite{Bedell2020}. Figure~\ref{fig1}(a) shows a schematic of device~1, comprising a JJ embedded in a superconducting PtSiGe loop (purple), with two gate electrodes (yellow) and a flux line (blue), which was absent for device~2. The PtSiGe contacts were fabricated via Pt lift-off and rapid thermal annealing~\cite{Tosato2023}. A $\mathrm{SiO_{x}}$ dielectric layer was deposited with atomic layer deposition and afterwards the gates and flux line were formed by lift-off of Ti/Al/Ti/Au. Appendix~A contains details of the fabrication procedure. The voltage $V_\mathrm{g}$ applied to the accumulation gate controlled the hole density in the Ge quantum well, which was insulating for $V_\mathrm{g}=0$. Unless specified differently, no dc voltage was applied to the side gate ($V_\mathrm{s}=0$). Instead, this gate was used for applying the rf drive tone for two-tone spectroscopy measurements via a bias-tee. The phase difference across the junction $\varphi$ was controlled with a magnetic flux $\Phi$, which we applied by injecting a current $I$ into the flux line (device~1) or into a superconducting coil in the vicinity of the sample (device~2). Due to the low kinetic inductance of the PtSiGe loops, we use the relation $\varphi=2\pi\Phi/\Phi_0$, where $\Phi_0$ is the superconducting flux quantum.

False-colored scanning electron micrographs of two exemplary junctions (top view) are shown in Figs.~\ref{fig1}(b) and (c). Device~1 consisted of a lithographically defined junction with length $L=350$~nm and width $W=100$~nm, whereas for device~2, $L=1.2~\mu$m and $W=150$~nm. The superconducting coherence length in Ge was estimated as $\xi_\mathrm{Ge}=\hbar v_\mathrm{F}/\pi\Delta_0\approx900~$nm, where $\Delta_0\approx60~\mu$eV (14.5~GHz) is the superconducting gap at $T=0$ and $v_\mathrm{F}\approx2.5\times10^5~\mathrm{m}~\mathrm{s}^{-1}$ is the Fermi velocity extracted from Hall bar measurements~\cite{Hinderling2024Ge} using an effective mass of $0.06m_e$~\cite{Lodari2019}. Since device~1 had a junction length smaller than the estimated coherence length, $L<\xi_\mathrm{Ge}$, this device was expected to conform to the short junction regime~\cite{Hinderling2024Ge}. For device~2, $L>\xi_\mathrm{Ge}$, placing this device in the long junction regime. The finite width of the devices implies that they could host multiple transverse conduction channels. 

Figure~\ref{fig1}(d) depicts an optical microscope image of a sample following flip-chip bonding and subsequent wire bonding. A high-resistivity Si chip (resonator chip) contained a superconducting feedline with three capacitively coupled $\lambda/4$ Nb coplanar waveguide resonators. During flip-chip bonding, the grounded ends of the Nb resonators on the resonator chip were aligned to the superconducting loops of the Ge devices on the device chip, resulting in inductive coupling. The two chips were connected via In bumps for electrical connections and mechanical stability, enabling application of dc and rf signals to the devices through wire bonds on the resonator chip. More details on the measurement circuit can be found in Refs.~\cite{Hinderling2023,Hinderling2024Ge}.

\begin{figure*}
	\includegraphics[width=0.95\textwidth]{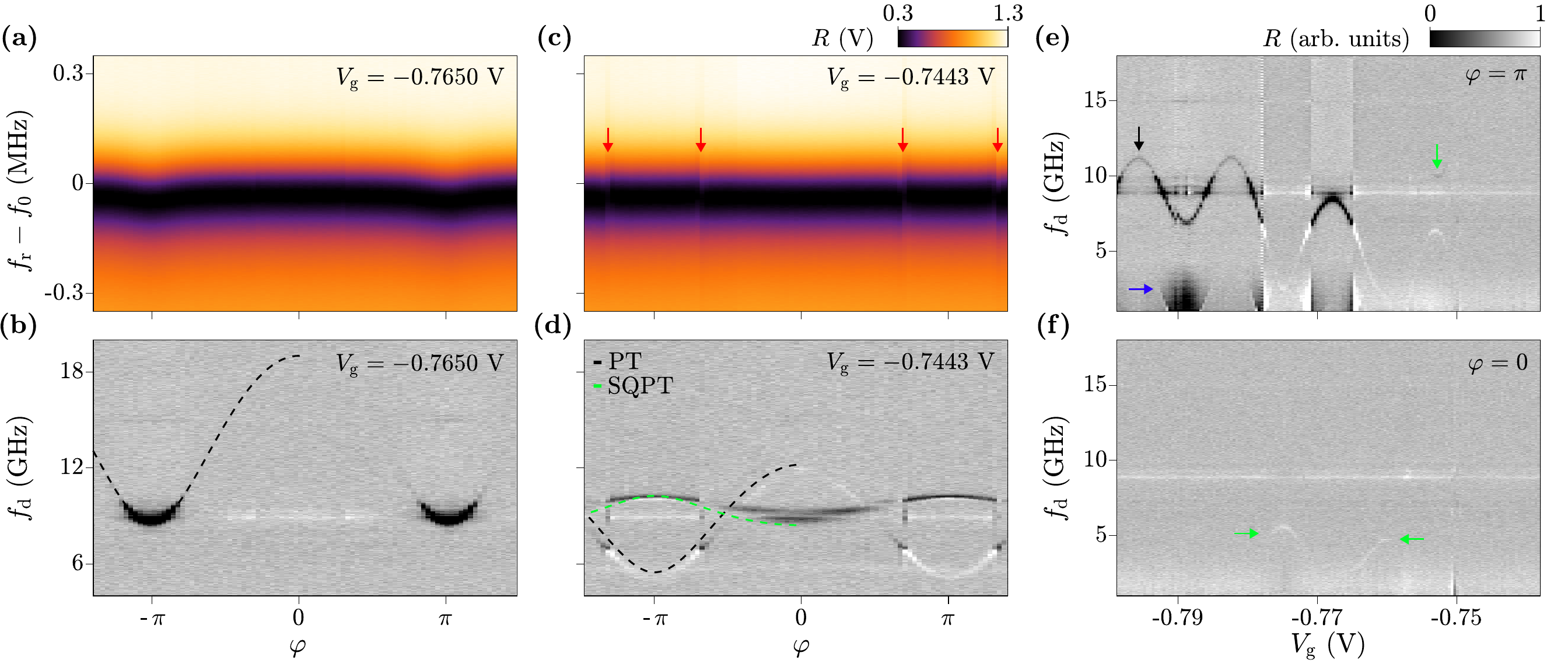}
	\caption{(a) Device~1: amplitude $R$ of the resonator transmission $S_{21}$ as a function of offset readout frequency $f_\mathrm{r}-f_0$ and phase $\varphi$ at gate voltage $V_\mathrm{g}=-0.7650$~V, with $f_\mathrm{0}=6.90115$~GHz. (b) Amplitude $R$ of the resonator transmission $S_{21}$ as a function of drive frequency $f_\mathrm{d}$ ($P_\mathrm{d}=-26$~dBm) and phase $\varphi$, measured together with (a). A single Andreev bound state (ABS) pair transition (PT) was observed. The black dashed line is a fit to the PT frequency $f_1=(2E_1)/h$ using Eq.~\ref{eq1} for $E_1$. (c) Same as (a), measured at $V_\mathrm{g}=-0.7443$~V. Anticrossings (red arrows) indicate ABS interaction with the resonator. (d) Same as (b), measured together with (c). Besides the PT, a single-quasiparticle transition (SQPT) was observed, indicating the presence of two ABSs. The black dashed line is a fit to the PT $f_\mathrm{1}=(2E_1)/h$ and the green dashed line corresponds to the SQPT with frequency $f=(E_2-E_1)/h$, where $E_1$ and $E_2$ are given by Eq.~\ref{eq1}. (e) Amplitude $R$ of the resonator transmission $S_{21}$ as a function of drive frequency $f_\mathrm{d}$ ($P_\mathrm{d}=-25$~dBm) and gate voltage $V_\mathrm{g}$, measured at $\varphi=\pi$. The PT (black arrow) was observed over nearly the full range of $V_\mathrm{g}$. The dark blue horizontal arrow indicates a replica of the PT. The SQPT is visible around $V_\mathrm{g}=-0.7550$~V (green arrow). (f) Same as (e), measured at $\varphi=0$. The SQPT appeared in a limited range of $V_\mathrm{g}$, indicated with green arrows.}
	\label{fig2}
\end{figure*}

For microwave measurements, a vector network analyzer (VNA) was used to apply a continuous readout tone with frequency $f_\mathrm{r}$ close to the resonance frequency of the resonator $f_\mathrm{res}$. The readout tone was transmitted through the feedline where it interacted with the resonator, and after amplification at various temperature stages of the dilution refrigerator the VNA detected the outgoing signal and determined the complex scattering parameter $S_\mathrm{21}$. Figures~\ref{fig1}(e) and (f) depict resonator traces, measured at $V_\mathrm{g}=0$, yielding the bare resonator frequencies $f_\mathrm{0}=6.90115$~GHz for device~1 and $f_\mathrm{0}=6.86731$~GHz for device~2. The internal ($Q_\mathrm{i}$) and loaded ($Q_\mathrm{L}$) quality factors of the resonators, obtained with the circle-fit method~\cite{Probst2015}, were $70000$ and $23000$, respectively, for device~1 and $355000$ and $39000$, respectively, for device~2. The large difference between the quality factors was attributed to the presence of the flux line for device~1, which was aligned parallel to the resonator and thus likely contributed to the microwave losses~\cite{Watanabe2009}. 

Two-tone microwave spectroscopy was performed by applying a drive tone with power $P_\mathrm{d}$ and frequency $f_\mathrm{d}$ to the side gate, in addition to the readout tone. Whenever the drive tone frequency was equal to a transition frequency $f_\mathrm{d}=\Delta E/h$ between different levels in the ABS energy spectrum, the frequency of the resonator acquired a dispersive shift and damping, enabling the detection of ABS transition frequencies~\cite{Romero2012,Janvier2015,Hays2018,Tosi2019,Park2020,Hays2020,Metzger2021,Chidambaram2022,Canadas2022,Zellekens2022,Fatemi2022,Hinderling2023,Wesdorp2024,Hinderling2024Ge,Elfeky2025}. Resonator compensation described in Appendix~C was done at each value of the gate voltage $V_\mathrm{g}$ or phase $\varphi$, such that two-tone spectroscopy was performed at a readout frequency with fixed offset from $f_\mathrm{res}$, which changed as a function of $V_\mathrm{g}$ and $\varphi$.

\section*{Microwave spectroscopy of device~1}

We first present microwave measurements on device 1, which had a length $L=350$~nm, smaller than the superconducting coherence length of Ge. Figure~\ref{fig2}(a) shows the amplitude $R$ of the resonator transmission as a function of the offset readout frequency $f_\mathrm{r}-f_0$ and the phase difference across the junction $\varphi$ at a gate voltage $V_\mathrm{g}=-0.7650$~V. The amplitude $R$ is given by $R=\sqrt{I^2+Q^2}$, where $I$ and $Q$ are the in-phase and quadrature components of the resonator response, respectively. The modulation of the resonance frequency of the resonator $f_\mathrm{res}$ with phase $\varphi$ was fit following Refs.~\cite{Haller2022,Hinderling2023}, yielding a critical current of $I_\mathrm{C}=3.1$~nA. Assuming a single ABS can carry up to $7$~nA of current \cite{Bagwell1992}, this is consistent with one or very few ABSs in the junction at this value of $V_\mathrm{g}$. 

\begin{figure*}
	\includegraphics[width=0.95\textwidth]{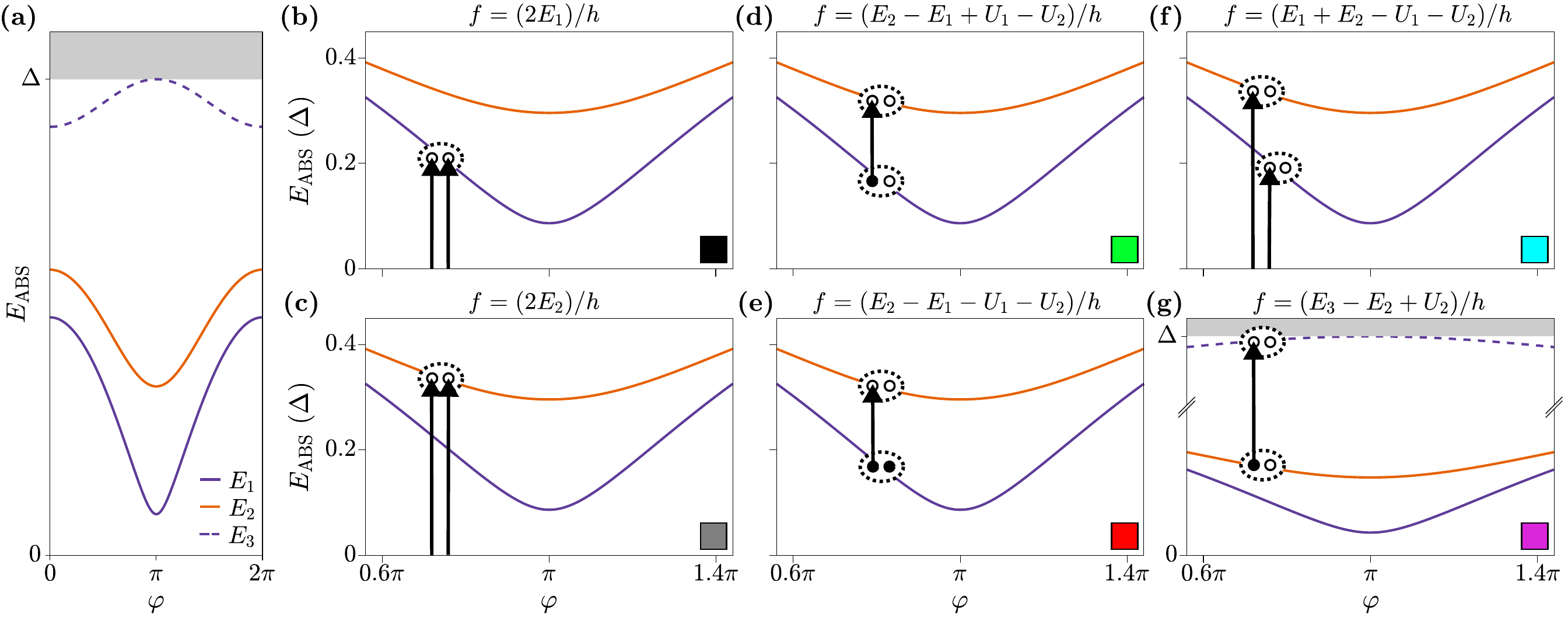}
	\caption{(a) Energy spectrum of a long junction hosting three spin-degenerate Andreev bound states (ABSs) in two conduction channels (purple, orange) with energies $E_1$ (purple, solid line), $E_2$ (orange, solid line) and $E_3$ (purple, dashed line) as a function of the phase difference across the junction $\varphi$, shown in the excitation picture. The continuum above the induced superconducting gap $\Delta$ is indicated with gray shading. Coulomb interaction manifests as an energy penalty $U_i$ given by Eq.~\ref{eq3} for states with even occupation. (b) Zoom-in of the energy spectrum in (a). The pair transition (PT) for the ABS with energy $E_1$ is indicated with arrows. The transition frequency is $f=(2E_1)/h$. (c) PT for the ABS with energy $E_2$, resulting in the transition frequency $f=(2E_2)/h$. (d) Single-quasiparticle transition (SQPT) from $E_1$ to $E_2$. The energy levels are shifted as a result of Coulomb interaction, yielding the transition frequency $f=(E_2 -E_1 +U_1 -U_2)/h$. (e) Same as (d), but with the first ABS initially in an excited state, resulting in $f=(E_2 -E_1 -U_1 -U_2)/h$. (f) Mixed PT with frequency $f=(E_1 + E_2 -U_1 -U_2)/h$. (g) SQPT from $E_2$ to $E_3$. Since Coulomb interaction is neglected for $E_3$ due to its close proximity to the continuum, the SQPT frequency is given by $f=(E_3-E_2+U_2)/h$.}
	\label{fig3}
\end{figure*}

A two-tone spectroscopy measurement is shown in Fig.~\ref{fig2}(b), acquired together with the single-tone measurement in Fig.~\ref{fig2}(a). Here, the amplitude $R$ is shown as a function of the drive tone frequency $f_\mathrm{d}$ and the phase difference across the junction $\varphi$. We observed a single transition, corresponding to a pair transition (PT) with frequency $f_1=(2E_1)/h$, as schematically shown in Fig.~\ref{fig3}(b). To account for the finite length of the junction, which results in detachment of the ABS from the continuum for a ballistic channel, the following dispersion relation~\cite{Kurilovich2021} was used for the ABS energy
\begin{equation}
	E_i(\varphi)=\pm \widetilde{\Delta}_i \sqrt{1-\tau_i \sin^2 (\varphi/2)},
	\label{eq1}
\end{equation}
\noindent where $\widetilde{\Delta}_i$ is the ABS amplitude and $\tau_i$ is the transmission of the state. Equation~\ref{eq1} is similar to the short-junction ABS dispersion relation~\cite{Beenakker1991_2,Bagwell1992}, except that $\widetilde{\Delta}_i<\Delta$, where $\Delta$ is the induced superconducting gap. Fitting the data in Fig.~\ref{fig2}(b) to Eq.~\ref{eq1} yielded $\widetilde{\Delta}/h\approx9.5$~GHz and $\tau\approx0.80$. The fit is plotted as a black dashed line in the left half of Fig.~\ref{fig2}(b). 

The ABS spectrum changed drastically when the hole density in the junction was decreased, as shown in Figs.~\ref{fig2}(c) and (d) at $V_\mathrm{g}=-0.7443$~V. The resonator trace in Fig.~\ref{fig2}(c) displays anticrossings around $\varphi=\pm\pi$ (red arrows), indicating that the ABS transition frequency $f_\mathrm{1}$ crossed the resonator frequency $f_\mathrm{res}$ at these values of the phase, resulting in ABS-resonator interaction~\cite{Chidambaram2022}. Due to the anticrossings, it was not possible to fit the modulation in $f_\mathrm{res}$ with the procedure used for Fig.~\ref{fig2}(a). The two-tone spectroscopy measurement in Fig.~\ref{fig2}(d) showed a PT with minima in the transition frequency at $\varphi=\pm\pi$ and a single-quasiparticle transition (SQPT), characterized by a minimum in the transition frequency at $\varphi=0$. The latter corresponds to a quasiparticle (QP) being excited from one ABS to another, resulting in a transition frequency $f=(E_2-E_1)/h$. Observation of this SQPT implies that the junction hosted more than one ABS at this value of $V_\mathrm{g}$. The SQPT was reproduced with the finite-length model of Eq.~\ref{eq1} by considering two ABSs with $\widetilde{\Delta}_1/h=0.42\Delta/h\approx6.1$~GHz ($\Delta=14.5$~GHz) and $\tau_1 =0.78$ for the first ABS, and $\widetilde{\Delta}_2/h=\Delta/h=14.5$~GHz and $\tau_2 =0.20$ for the second ABS. The high value of $\widetilde{\Delta}_2$ and low value of $\tau_2$ indicate that the energy of the second ABS remained close to the superconducting gap $\Delta$, with weak dispersion. The PT associated to the second ABS with frequency $f_2=(2E_2)/h$ is outside the spectral range of Fig.~\ref{fig2}(d). Therefore, the second ABS was only observed indirectly via the SQPT. Note that sign inversions of the shift in $R$ occurred whenever the ABS transition frequency crossed the resonator frequency at $f_\mathrm{d}=f_\mathrm{res}\approx6.90115$~GHz, resulting from a change in the direction of the resonator dispersive shift. Moreover, the dispersive shift is opposite for PTs and SQPTs~\cite{Metzger2021}.

The evolution of the ABS transition frequencies with the gate voltage $V_\mathrm{g}$ is depicted in Figs.~\ref{fig2}(e) and (f) at phases $\varphi=\pi$ and $\varphi=0$, respectively. These two-tone spectroscopy measurements were acquired together, where for each value of $V_\mathrm{g}$, resonator compensation and a two-tone spectroscopy measurement were performed at both $\varphi=\pi$ and $\varphi=0$ (Appendix~D). The PT corresponding to the first ABS, indicated with a black arrow in Fig.~\ref{fig2}(e), was observed over nearly the full range of $V_\mathrm{g}$ at $\varphi=\pi$, with replicas appearing at $6.9$~GHz below the transition frequency, indicated with a dark blue horizontal arrow. Such replicas are commonly observed in two-tone spectroscopy measurements and are attributed to multi-photon processes involving resonator photons with energy $E=hf_\mathrm{res}$ \cite{Zellekens2022,Hinderling2023}. In addition, the SQPT was observed around $V_\mathrm{g}=-0.7550$~V [green arrow in Fig.~\ref{fig2}(e)]. For the measurement performed at $\varphi=0$ in Fig.~\ref{fig2}(f), the PT with $f_1(\varphi=0)>f_1(\varphi=\pi)$ was not observed. Instead, the SQPT is visible in a finite range of $V_\mathrm{g}$, indicated with green arrows. The appearance of the SQPT for a limited range of $V_\mathrm{g}$ values might be due to opening and closing of the second conduction channel and oscillations of the second ABS, similar to those of the first ABS in Fig.~\ref{fig2}(e). Mesoscopic fluctuations of the ABS energy as a function of gate voltage have also been reported for JJs realized in nanowires and 2DEGs~\cite{Tosi2019,Zellekens2022,Hinderling2023,Wesdorp2024,Sahu2024,Elfeky2025}. Due to gate-voltage induced hysteresis, which was measured in the same material~\cite{Massai2024}, the transition frequencies at $\varphi=0$ and $\varphi=\pi$ in Figs.~\ref{fig2}(a-d) deviate from those in Figs.~\ref{fig2}(e) and (f) at the corresponding values of $V_\mathrm{g}$. The gate dependence of an additional device, with similar dimensions to device~1, is presented in Appendix~E. 

In summary, device~1 showed finite-length effects despite the junction length being shorter than the estimated superconducting coherence length of Ge. Moreover, readout improvements enabled the observation of an SQPT, previously not detected for similar junctions~\cite{Hinderling2024Ge}. These transitions are particularly interesting for studies of spin-orbit interaction in long junctions~\cite{Tosi2019,Hays2020,Hays2021,Canadas2022,Wesdorp2024}.

\begin{figure*}
	\includegraphics[width=0.95\textwidth]{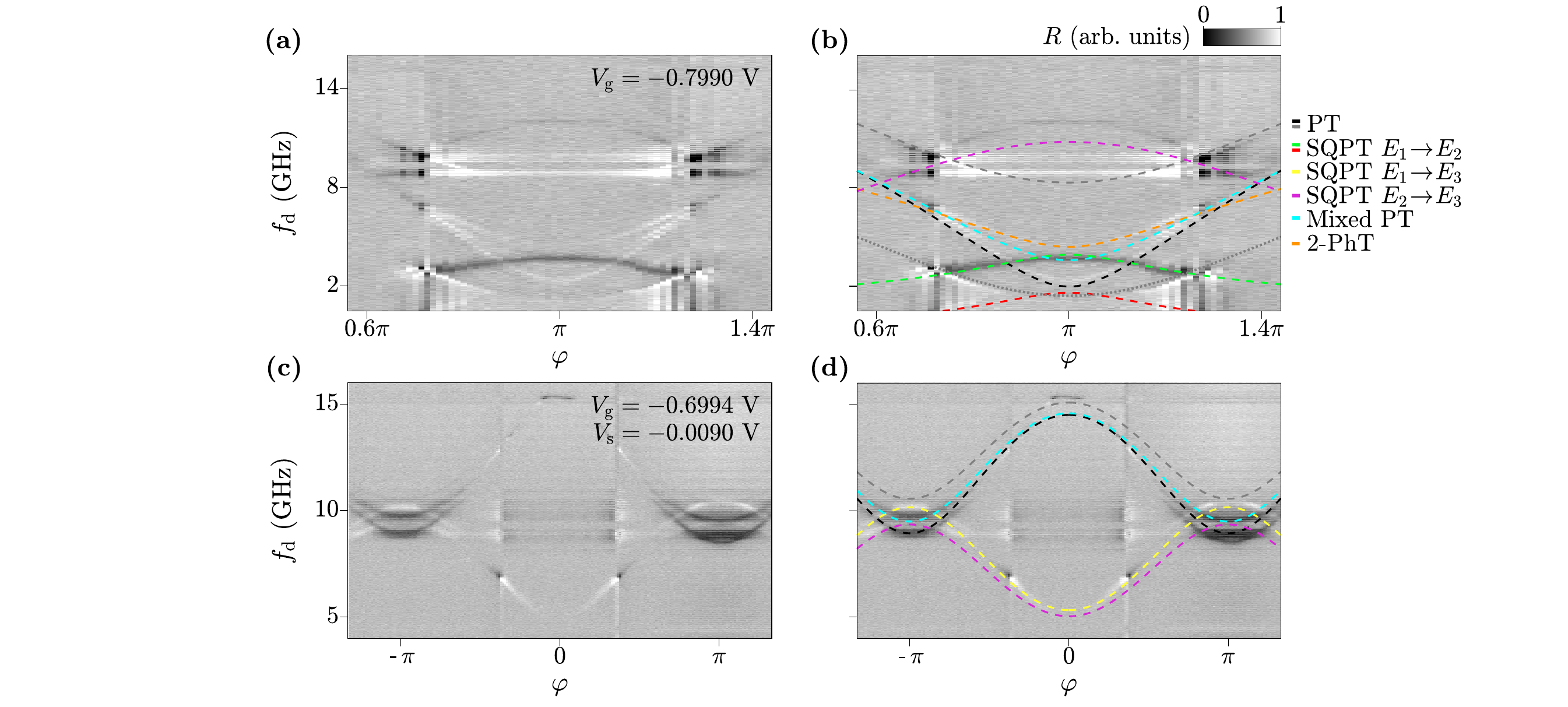}
	\caption{(a) Device~2: amplitude $R$ of the resonator transmission $S_{21}$ as a function of drive frequency $f_\mathrm{d}$ ($P_\mathrm{d}=-28$~dBm) and phase $\varphi$, measured at $V_\mathrm{g}=-0.7990$~V. Multiple pair transitions (PTs) and single-quasiparticle transitions (SQPTs) are visible around $\varphi=\pi$. (b) Same as (a), but overlaid with ABS transitions predicted by the long junction model with three ABSs and including Coulomb interaction. The black and gray dashed lines are PTs, schematically shown in Figs.~\ref{fig3}(b) and (c). The gray dotted line is a replica of the gray PT, shifted downward by the resonator frequency $f_\mathrm{res}\approx6.86731$~GHz. The green, red and purple dashed lines are SQPTs corresponding to Figs.~\ref{fig3}(d), (e) and (g), respectively. The cyan line is a mixed PT [Fig.~\ref{fig3}(f)] and the orange line is a two-photon transition (2-PhT) involving the lowest-energy PT and a resonator photon with energy $E=hf_\mathrm{res}$. (c) Same as (a), but measured over a wider range of $\varphi$ and at $V_\mathrm{g}=-0.6994$~V and $V_\mathrm{s}=-0.0090$~V ($P_\mathrm{d}=-35$~dBm). At this gate voltage configuration, several PTs and SQPTs were observed. (d) Same as (c), overlaid with ABS transitions predicted by the model. The transitions are colored according to the legend in (b). Here, the yellow dashed line is a SQPT from the lowest-energy ABS $E_1$ to the highest-energy ABS $E_3$.}
	\label{fig4}
\end{figure*}

\section*{Coulomb interaction in a long junction}

For junctions with length $L\sim\xi$, QPs have a finite dwell time in the junction region and therefore interactions affect the ABS physics. In particular, Coulomb interaction leads to energy shifts in the ABS spectrum~\cite{Kurilovich2021,Fatemi2022,Canadas2022,Kurilovich2024}. To reproduce experimentally observed ABS transitions presented in the next section, we consider a long junction with two conduction channels, each of which can host up to two ABSs. The corresponding ABS transitions as well as the effects of Coulomb interaction are discussed here.

Figure~\ref{fig3}(a) schematically depicts the energy spectrum of a finite-length junction with two conduction channels hosting three spin-degenerate ABSs, $E_1$ (solid line) and $E_3$ (dashed line) in one conduction channel (purple) and $E_2$ (solid line) in the second conduction channel (orange). The energy of the first ABS, $E_1$, is smaller than the energy of the second ABS, $E_2$. Since the energy spectrum is symmetric around zero energy, the excitation picture is used here, for which only the positive part of the energy spectrum is depicted. The ABS energies $E_i$ are given by Eq.~\ref{eq1}. As a result of Coulomb interaction in the junction, the energies of the many-body states shift depending on the population of the ABSs. Since the charge of a QP depends on its energy, the phase dependence of the Coulomb interaction energy $U_i$ for an ABS with energy $E_i$ from Eq.~\ref{eq1} is estimated as~\cite{Kurilovich2024}
\begin{equation}
	U_i(\varphi)=\begin{cases}
		u_i \left[\Delta^2 - E_i(\varphi)^2\right], & n=0,2\\
		0, & n=1
	\end{cases}
	\label{eq3}
\end{equation}
\noindent where $n$ is the number of QPs occupying an ABS and $u_i$ is a constant, defined by the microscopic details of the corresponding state and the extension of the wave function into the superconducting leads, see Appendix~F. Each ABS can host up to two QPs. For states occupied by a single QP, particle-hole symmetry implies that $U_i (\varphi)=0$. Instead, states with even occupation, i.e., zero or two QPs, are shifted to higher energy by $+U_i (\varphi)$~\cite{Kurilovich2021,Fatemi2022}. Consequently, the frequencies of certain ABS transitions are affected by Coulomb interaction. In Figs.~\ref{fig3}(b-g), different transitions involving the two lowest-energy ABSs $E_1$ and $E_2$ (b-f) and the third ABS $E_3$ (g) are schematically illustrated. Below, we discuss the effects of Coulomb interaction on these transitions. We neglect Coulomb interaction for $E_3$, since the energy of this state is close to the continuum edge $\Delta$. More details are reported in Appendix~F.

During a PT, two QPs are excited from the ground state to the ABS. Figure~\ref{fig3}(b) shows the PT for the first ABS with energy $E_1$ (purple), where $E_1$ is given by Eq.~\ref{eq1}. Before and after the PT, the energy of the ABS is shifted by $+U_1$, since the occupation of the state is even. Therefore, the energy change due to the PT, which determines the transition frequency is given by $\Delta E=2(E_1+U_1)-2U_1)=2E_1$, yielding $f=(2E_1)/h$ for the PT. Accordingly, the frequencies of all PTs are not affected by Coulomb interaction. The PT for the second ABS with energy $E_2$ (orange) is shown in Fig.~\ref{fig3}(c) with transition frequency $f=(2E_2)/h$.

Coulomb interaction modifies the frequency of the SQPT from $E_1$ to $E_2$ in Fig.~\ref{fig3}(d). Before the SQPT, the energy is given by $E_1+U_2$, because a single QP occupies $E_1$, resulting in no energy shift for this level, and the unoccupied level $E_2$ is shifted by $+U_2$. After the QP is excited to $E_2$, the energy is given by $E_2+U_1$, as the first ABS is now unoccupied, giving rise to an energy shift of $+U_1$, while the second ABS hosting one QP has no energy shift. Consequently, the frequency of the SQPT in Fig.~\ref{fig3}(d) is given by $f=(E_2-E_1+U_1-U_2)/h$. If $E_1$ is initially in an excited state with full occupation as shown in Fig.~\ref{fig3}(e), before the SQPT both $E_1$ and $E_2$ have even occupation and after the SQPT both states are occupied by a single QP, resulting in an SQPT frequency $f=[E_2-(E_1+U_1+U_2)]/h$. Hence, an important consequence of Coulomb interaction is that the degeneracy between the frequencies of the SQPTs in Figs.~\ref{fig3}(d) and (e) is lifted. The frequency of the mixed pair transition (mixed PT) in Fig.~\ref{fig3}(f) is also modified by Coulomb interaction. Here, $f=(E_1+E_2-U_1-U_2)/h$ since the energy before the transition is $U_1+U_2$, while the energy after the transition is $E_1+E_2$. Finally, the SQPT to the third ABS shown in Fig.~\ref{fig3}(g) has a transition frequency given by $f=(E_3-E_2+U_2)/h$, since we neglect Coulomb interaction for $E_3$.

\section*{Microwave spectroscopy of device~2}

Device~2 had a junction length $L=1.2~\mu$m, larger than the superconducting coherence length of Ge. Two-tone spectroscopy measurements in Fig.~\ref{fig4} revealed the presence of multiple ABSs in the junction. Figure~\ref{fig4}(a) depicts the amplitude $R$ of the resonator transmission $S_{21}$ as a function of the phase difference across the junction $\varphi$ and the drive tone frequency $f_\mathrm{d}$, at $V_\mathrm{g}=-0.7990$~V. Multiple transitions are visible in the ABS spectrum, with varying linewidth and intensity. Similar to device~1, sign inversions of the shift in $R$ occur whenever a transition frequency crosses the resonator frequency $f_\mathrm{res}$, and the shift is opposite for PTs with minima at $\varphi=\pi$ and SQPTs with maxima at $\varphi=\pi$. The presence of multiple SQPTs implied that the junction hosted several ABSs for this gate configuration. The observations were explained by a model considering three ABSs in the junction with Coulomb interaction, introduced in the previous section. The ABS energies $E_1$ and $E_2$ are given by Eq.~\ref{eq1} with $\widetilde{\Delta}_1/h=0.48\Delta/h\approx7.0$~GHz and $\tau_1 =0.98$, and $\widetilde{\Delta}_2/h=0.54\Delta/h\approx7.8$~GHz and $\tau_2 =0.72$, while $E_3$ was approximated using a cosine function, since this ABS was close to the continuum edge with weak dispersion and therefore it was only observed indirectly (see Appendix~F). The PTs to $E_1$ and $E_2$ are depicted as black and gray dashed lines in Fig.~\ref{fig4}(b), matching the color of the square symbols in Figs.~\ref{fig3}(b) and (c). Moreover, a replica of $E_2$ was observed, shifted downwards by the resonator frequency to $f_\mathrm{d}=f_2-6.9$~GHz. This replica is indicated with a gray dotted line in Fig.~\ref{fig4}(b). The SQPTs between $E_1$ and $E_2$, indicated with green and red dashed lines, respectively, correspond to the transitions in Figs.~\ref{fig3}(d) and (e), while the SQPT from $E_2$ to $E_3$, marked with a purple dashed line, corresponds to Fig.~\ref{fig3}(g). The mixed PT, schematically shown in Fig.~\ref{fig3}(f), is indicated with a cyan dashed line in Fig.~\ref{fig4}(b). Moreover, the transition at slightly higher energy than the mixed PT can likely be identified as a two-photon transition (2-PhT) \cite{Janvier2015} involving the PT to $E_1$ and a resonator photon with energy $hf_\mathrm{res}$, which yields the transition frequency $f=(2E_1+hf_\mathrm{res})/2h$, indicated with an orange dashed line in Fig.~\ref{fig4}(b). Presumably, this 2-PhT was observed because the frequency of this PT crossed the resonator frequency, as evidenced by the single-tone spectroscopy measurements in Appendix~G.

Changing the gate voltage modified the ABS spectrum significantly. In Fig.~\ref{fig4}(c), we show another two-tone spectroscopy measurement of device~2, at a gate voltage $V_\mathrm{g}=-0.6994$~V, with the side-gate voltage at $V_\mathrm{s}=-0.0090$~V. Here, besides two PTs with minima at $\varphi=\pm\pi$, two SQPTs with distinct maxima at $\varphi=\pm\pi$ were observed, indicating the presence of a third ABS also for this gate voltage configuration. The ABS energies $E_1$ and $E_2$ were found with Eq.~\ref{eq1} using $\widetilde{\Delta}_1/h=0.50\Delta\approx7.3$~GHz and $\tau_1=0.62$, and $\widetilde{\Delta}_2/h=0.52\Delta\approx7.5$~GHz and $\tau_2 =0.51$. Again, the ABS with energy $E_3$ resided close to the continuum edge and thus it was approximated using a cosine function (Appendix~F). Consequently, the PT to $E_3$ was outside the spectral range of the measurement. The PTs to $E_1$ and $E_2$ are indicated with black and gray dashed lines in Fig.~\ref{fig4}(d). The minima of the PTs were slightly different at $\varphi=-\pi$ and $\varphi=\pi$, which we attributed to gate voltage instabilities in the device, see also Appendix~H. Due to the small energy difference between $E_1$ and $E_2$, SQPTs between these states, shown in Figs.~\ref{fig3}(d) and (e), were not observed here. Instead, the SQPTs from $E_1$ to $E_3$ (yellow dashed line) and $E_2$ to $E_3$ (purple dashed line) matched the experimental data. The frequency difference between these SQPTs was enhanced by the Coulomb interaction, which was stronger for the first ABS. This is evident from Eq.~\ref{eq3} where the Coulomb interaction energy $U_i$ is larger for lower $E_i$. The other transition in Fig.~\ref{fig4}(c) with minima at $\varphi=\pm \pi$ corresponds to the mixed PT between $E_1$ and $E_2$ and is indicated with a cyan dashed line in Fig.~\ref{fig4}(d). Finally, we note that fewer transitions are visible in Fig.~\ref{fig4}(c) than in Fig.~\ref{fig4}(a), which could be related to the different gate voltages or the lower drive power $P_\mathrm{d}$ used for Fig.~\ref{fig4}(d).

The experimental data are in good agreement with the model, apart from the SQPT from $E_1$ to $E_3$ in Fig.~\ref{fig4}(b) (purple dashed line), for which the model predicts slightly lower transition frequencies. To achieve better quantitative agreement between the measured and predicted transition frequencies, higher-order terms in the Coulomb interaction, as well as the exchange interaction~\cite{Canadas2022,Wesdorp2024} would need to be included in the model. 

We note that signatures of spin splitting in ABSs, reported for long junctions in InAs nanowires~\cite{Tosi2019,Hays2020,Hays2021,Canadas2022,Zellekens2022,Wesdorp2024} were not observed in our devices. This is surprising as device~2 conformed to the long junction regime and showed several SQPT and mixed PT transitions, which generally evidence spin-orbit effects. We therefore conclude that spin-orbit interaction in our devices is weak, which is presumably related to strain in the Ge quantum well~\cite{Bosco2021,Adelsberger2022} or the junction geometry~\cite{Hoffman2025}, both of which suppress spin-orbit interaction. Furthermore, two-dimensional hole gases have a cubic Rashba spin-orbit interaction, where spin splitting for low carrier density is expected to be small~\cite{Luethi2023}. Spin-orbit interaction can likely be strengthened by engineering the material stack. Strain in the quantum well might be reduced by optimizing the stoichiometry of the SiGe barriers and buffer layers. In addition, it was recently shown that spin-orbit interaction was enhanced in an unstrained Ge-SiGe heterojunction~\cite{Costa2025}.

Several transitions are predicted by the model with frequencies outside the spectral range of the measurements. Not only was the setup limited by the frequency range of the VNA (26~GHz), the intensity of transitions decreased as the detuning from the resonator increased. To measure transitions at high frequencies, the coupling between the resonator and the ABS would need to be improved. This is a limitation of the flip-chip approach, where the air gap between the resonator and device chips limits the strength of the inductive coupling to several MHz~\cite{Hinderling2024Ge}. Nonetheless, future work could focus on optimizing the resonator-ABS coupling strength by decreasing the air gap between the flip-chip bonded chips, modifying the resonator frequency or investigating different types of resonators~\cite{Shvetsov2025}. 

\section*{Conclusion}\label{sec13}

In summary, we performed microwave spectroscopy measurements of Andreev bound states in Josephson junctions shorter and longer than the superconducting coherence length, realized in proximitized Ge quantum wells. Flip-chip bonding enabled inductive coupling between superconducting PtSiGe loops and Nb coplanar waveguide resonators, resulting in high quality factors. Two-tone microwave spectroscopy measurements of a $350$~nm long junction revealed a single-quasiparticle transition in a limited range of gate voltage values, previously not observed in similar devices~\cite{Hinderling2024Ge}. We developed a model that was able to reproduce the experimental data by taking the finite length of the junction into account. Our model revealed the presence of a second ABS whose energy was outside the spectral range of the measurements. A device consisting of a junction with length $1.2~\mu$m was also investigated. Several ABS transitions were measured with two-tone spectroscopy, indicative of multiple interacting ABSs in the junction. The experimentally observed transitions were reproduced with a model comprising three ABSs in two conducting channels with finite Coulomb interaction. Such a model provided good qualitative fits to the observations for multiple gate configurations.

We demonstrated the importance of including finite-length effects and Coulomb interaction in models describing ABSs in intermediate and long junctions. Next steps for understanding the ABS physics in proximitized Ge quantum wells include theoretical investigation of spin-orbit interaction in Ge-based Josephson junctions and experimental studies of dynamic processes, such as quasiparticle poisoning and parity switching, in devices where Coulomb interaction is prominent.

\section*{Acknowledgments}
We thank the Cleanroom Operations Team of the Binnig and Rohrer Nanotechnology Center (BRNC) for their help and support. We thank S.~Bosco, B.~van~Heck, V.~Coppini, J.~Klinovaja, V.~Kozin, M.~Pita-Vidal, D.~Z.~Haxell and H.~Riel for useful discussions. J.~C.~C. thanks the Spanish Ministry of Science and Innovation (Grant~No.~PID2020-114880GB-I00) for financial support and the Deutsche Forschungsgemeinschaft (DFG; German Research Foundation) via SFB~1432 for sponsoring his stay at the University of Konstanz as a Mercator Fellow. D.~C.~O., A.~E.~S. and W.~B. acknowledge support by the Deutsche Forschungsgemeinschaft (DFG; German Research Foundation) via SFB 1432 (Project No. 425217212) and by the Excellence Strategy of the University of Konstanz via a Blue Sky project. F.~N. acknowledges support from the Swiss National Science Foundation (Grant~No.~200021~201082). 

\section*{Data availability}
Data presented in this work is available on Zenodo at [url inserted when published].

\section*{APPENDIX A: DEVICE FABRICATION}

The Josephson junction devices studied here were realized using a SiGe heterostructure grown with reduced-pressure chemical vapor deposition~\cite{Bedell2020}. The material stack consisted of a Ge layer directly grown on a Si wafer, followed by a SiGe layer with an increasing Si content, up to Si${}_{0.2}$Ge${}_{0.8}$. This stoichiometry was used for the buffer layers surrounding the 20-nm thick Ge quantum well, located 48~nm below the surface. More details on the SiGe heterostructure can be found in Refs.~\cite{Bedell2020,Massai2024}. The superconducting loops were realized by electron beam lithography, evaporation and lift-off of Pt, followed by rapid thermal annealing at 350\textdegree C, resulting in superconducting germanosilicide (PtSiGe), directly contacting the Ge quantum well~\cite{Hinderling2024Ge}. Prior to the Pt evaporation, the native oxide was removed by a short submersion in hydrofluoric acid (HF). To isolate the gates from the heterostructure, a 12~nm thick layer of $\mathrm{SiO_{x}}$ was deposited using atomic layer deposition. Then the inner gates, close to the active areas of the devices, were fabricated by evaporation and lift-off of Ti/Al/Ti/Au (5~nm/70~nm/5~nm/40~nm). Afterwards, the outer gates and flux line, consisting of Ti/Al/Ti/Au (5~nm/300~nm/5~nm/50~nm), were evaporated. For device~2, a single gate layer was used to define the gates, made of Ti/Al/Ti/Au (5~nm/70~nm/5~nm/40~nm). Finally, the chip was diced into a 3~mm by 3~mm piece for flip-chip bonding to a resonator chip. 

The resonators were fabricated on a high-resistivity ($\rho > 10~\mathrm{k}\Omega$~cm) intrinsic Si wafer. After removal of the native oxide using HF, a 200~nm thick layer of Nb was sputtered on the surface of the wafer. Using an Al${}_2$O${}_3$ hard mask and inductively coupled plasma
reactive-ion etching with Cl${}_2$/Ar, the resonators, feedline and DC control lines, as well as ground plane holes, were patterned into the Nb layer. For flip-chip bonding, In bumps were realized with evaporation and lift-off. Prior to the In evaporation, the native oxide was removed using HF.

\begin{figure*}
	\includegraphics[width=1.0\textwidth]{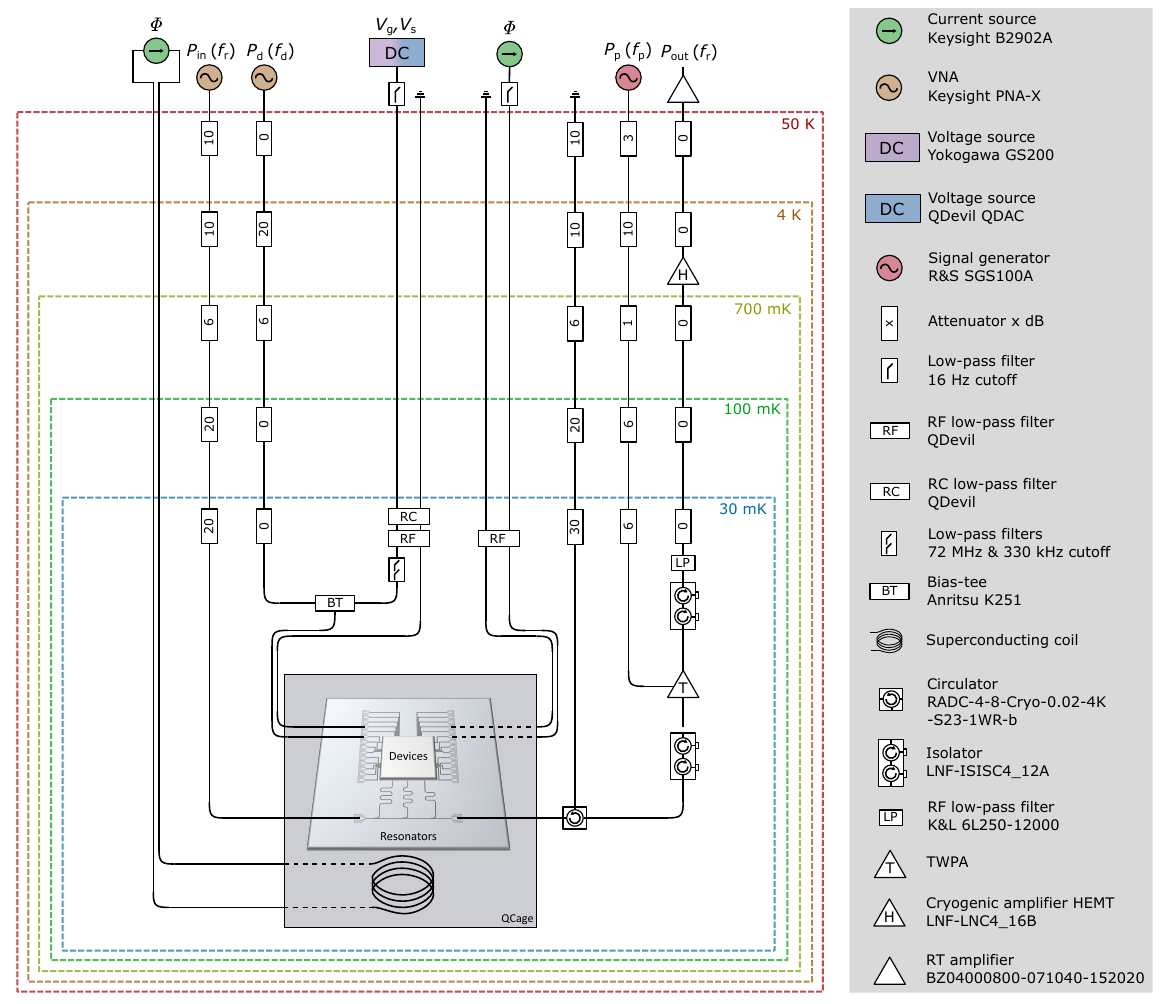}
	\caption{Schematic representation of the flip-chip bonded device in the QCage.24 sample holder, the dilution refrigerator used to perform the measurements, and the electronic setup.}
	\label{fig5}
\end{figure*}

\section*{APPENDIX B: PACKAGING AND MEASUREMENT SETUP}

The device and resonator chips were flip-chip bonded with a Karl Suss FC 150 flip-chip bonder. Alignment markers on both chips facilitated correct positioning of the device chip with respect to the resonator chip, with the superconducting device loops aligned to the grounded ends of the $\lambda/4$~coplanar waveguide resonators. Afterwards, the resonator chip was clamped in the QDevil QCage.24 sample holder with an integrated superconducting coil for magnetic flux biasing. Microwave and dc lines on the sample holder were connected to corresponding bonding pads on the resonator chip via wire bonding with Al bond wires. Moreover, Al air bridges across the resonators and feedline connected different areas of the ground plane, to establish a uniform ground potential across the resonator chip.

Measurements were performed in a BlueFors BF~LD~400 cryogen-free dilution refrigerator with a mixing chamber base temperature of 30~mK. Figure~\ref{fig5} schematically illustrates the wiring of the dilution refrigerator. A Keysight PNA-X vector network analyzer (VNA) was used to apply a readout tone with power $P_\mathrm{in}$ and frequency $f_\mathrm{r}$. Two-tone spectroscopy measurements were performed by applying a continuous drive tone of power $P_\mathrm{d}$ and frequency $f_\mathrm{d}$ with the same VNA. The readout and drive signals were attenuated by 66~dB and 26~dB, respectively. After propagating through the feedline, a circulator and a dual isolator, the readout signal was amplified by a traveling wave parametric amplifier (TWPA) with 20~dB gain at the mixing-chamber temperature stage of the dilution refrigerator. The TWPA was pumped by a continuous pump tone supplied by an R\&S SGS100A signal generator and attenuated by 26~dB in the dilution refrigerator. Then, the readout signal passed through another dual isolator and a low-pass filter, before amplification by a cryogenic high electron mobility transistor (HEMT) amplifier with 36~dB gain, installed at the 4~K stage, and a room-temperature amplifier with 44~dB gain. The amplified signal was detected by the VNA, at the port labeled $P_\mathrm{out}$ in Fig.~\ref{fig5}. The gate voltages $V_\mathrm{g}$ and $V_\mathrm{s}$ were applied by a Yokogawa GS200 dc voltage source (device~1) and a QDevil digital-to-analog converter dc voltage source (device~2). The dc signals were filtered by a home-made low-pass filter at room temperature and QDevil RC and rf filters at the mixing chamber stage of the dilution refrigerator before passing through bias-tees to combine dc and rf drive signals. For device~1, the magnetic flux threading the superconducting device loop was supplied by an on-chip flux line, which was current-biased by a Keysight B2902A SMU. The current passed through a NbTi superconducting line, filtered by a QDevil rf filter at the mixing chamber stage of the dilution refrigerator. For device~2, the magnetic flux was controlled by applying a current with the Keysight B2902A SMU to a superconducting coil integrated in the QCage.24 sample holder. The sample space was additionally shielded from external magnetic fields by a home-made magnetic shield comprising a $\mu$-metal and superconducting sheet.

\section*{APPENDIX C: RESONATOR COMPENSATION}

The two-tone spectroscopy measurements shown in Figs.~\ref{fig2} and \ref{fig4} were acquired using resonator compensation. Each time the slow axis variable, i.e., gate voltage or magnetic flux (phase) was changed, the drive tone was switched off and a resonator trace was measured, for which the readout frequency $f_\mathrm{r}$ was swept around the resonance frequency of the resonator and the complex scattering parameter $S_{21}$ was recorded to precisely determine $f_\mathrm{res}$, the frequency of the minimum in the magnitude $\abs{S_{21}}$. Then, the readout frequency $f_\mathrm{r}$ was fixed at $f_\mathrm{r}=f_\mathrm{res}+50$~kHz, on the slope of the resonator trace, where the sensitivity of $\abs{S_{21}}$ to shifts in $f_\mathrm{res}$ was high. After fixing the readout frequency, the drive tone with power $P_\mathrm{d}$ was applied to the device and a two-tone spectroscopy measurement was performed by recording $S_{21}$ at the aforementioned fixed readout frequency $f_\mathrm{r}$ while sweeping the drive frequency $f_\mathrm{d}$. For these measurements, the readout power was fixed at $P_\mathrm{in}=-40$~dBm for device~1 (Fig.~\ref{fig2}) and $P_\mathrm{in}=-42$~dBm for device~2 (Fig.~\ref{fig4}). Two-tone spectroscopy measurements were acquired with an integration time per point of $100~\mu$s and 500~averages. 

During post-processing of the two-tone spectroscopy data, the background was removed by subtracting the median along the drive frequency axis for each value of the gate voltage or phase.

\section*{APPENDIX D: DEVICE 1 GATE DEPENDENCE}

\begin{figure}[H]
	\includegraphics[width=0.45\textwidth]{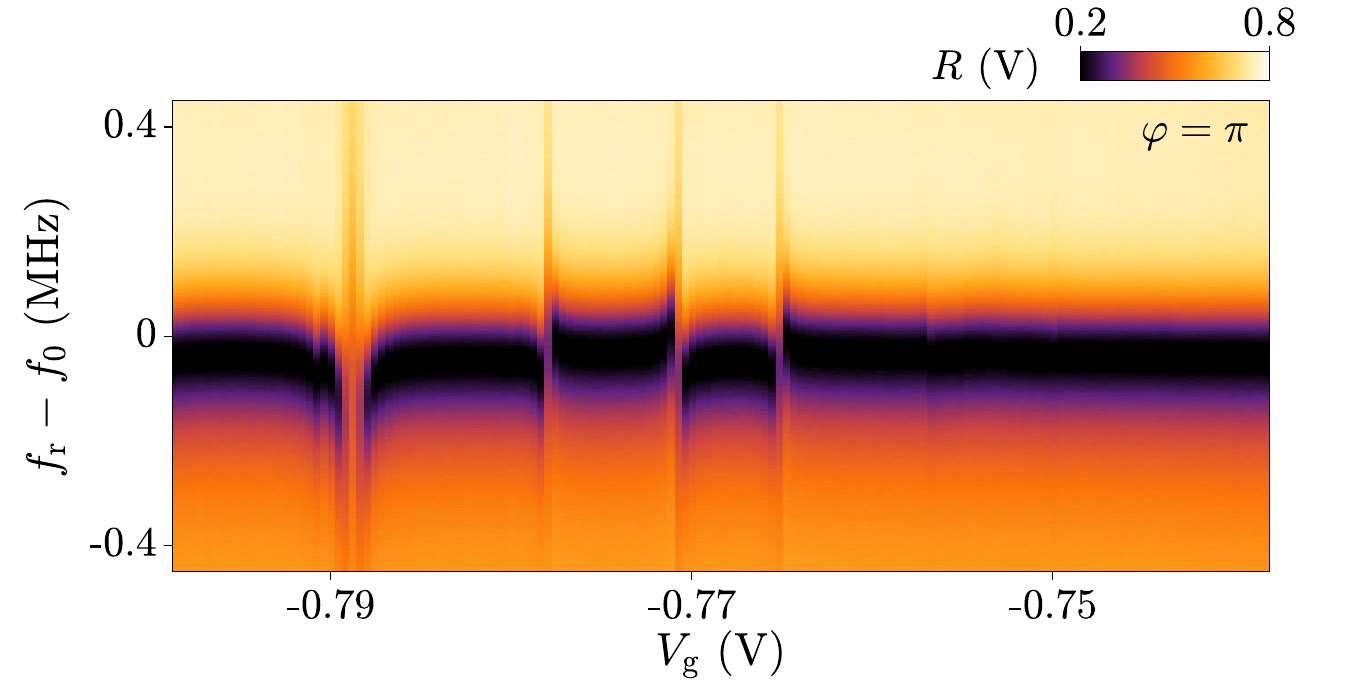}
	\caption{Device~1: amplitude $R$ of the resonator transmission $S_{21}$ as a function of offset readout frequency $f_\mathrm{r}-f_0$ and gate voltage $V_\mathrm{g}$, with $f_\mathrm{0}=6.90115$~GHz. The measurement was performed at $\varphi=\pi$ together with the two-tone spectroscopy measurement in Fig.~\ref{fig2}(e).}
	\label{fig7}
\end{figure}

The two-tone spectroscopy measurements as a function of gate voltage at $\varphi=\pi$ in Fig.~\ref{fig2}(e) and $\varphi=0$ in Fig.~\ref{fig2}(f) were acquired together with resonator traces at each value of the gate voltage $V_\mathrm{g}$ to perform resonator compensation. In Fig.~\ref{fig7}, the resonator traces at $\varphi=\pi$ are combined, yielding the amplitude $R$ of the resonator transmission $S_{21}$ as a function of the offset readout frequency $f_\mathrm{r}-f_0$ and the gate voltage $V_\mathrm{g}$. Anticrossings indicate interaction between the Andreev bound state (ABS) and the resonator, when the ABS transition frequency $f_1$ was equal to the resonance frequency of the resonator $f_\mathrm{res}$. When $f_1$ was smaller than $f_\mathrm{res}$, which is evident from the two-tone spectroscopy measurement in Fig.~\ref{fig2}(e), the resonance frequency of the resonator $f_\mathrm{res}$ was shifted to higher frequencies than when $f_1> f_\mathrm{res}$.

\section*{APPENDIX E: DEVICE 3 GATE DEPENDENCE}

\begin{figure}[H]
	\includegraphics[width=0.45\textwidth]{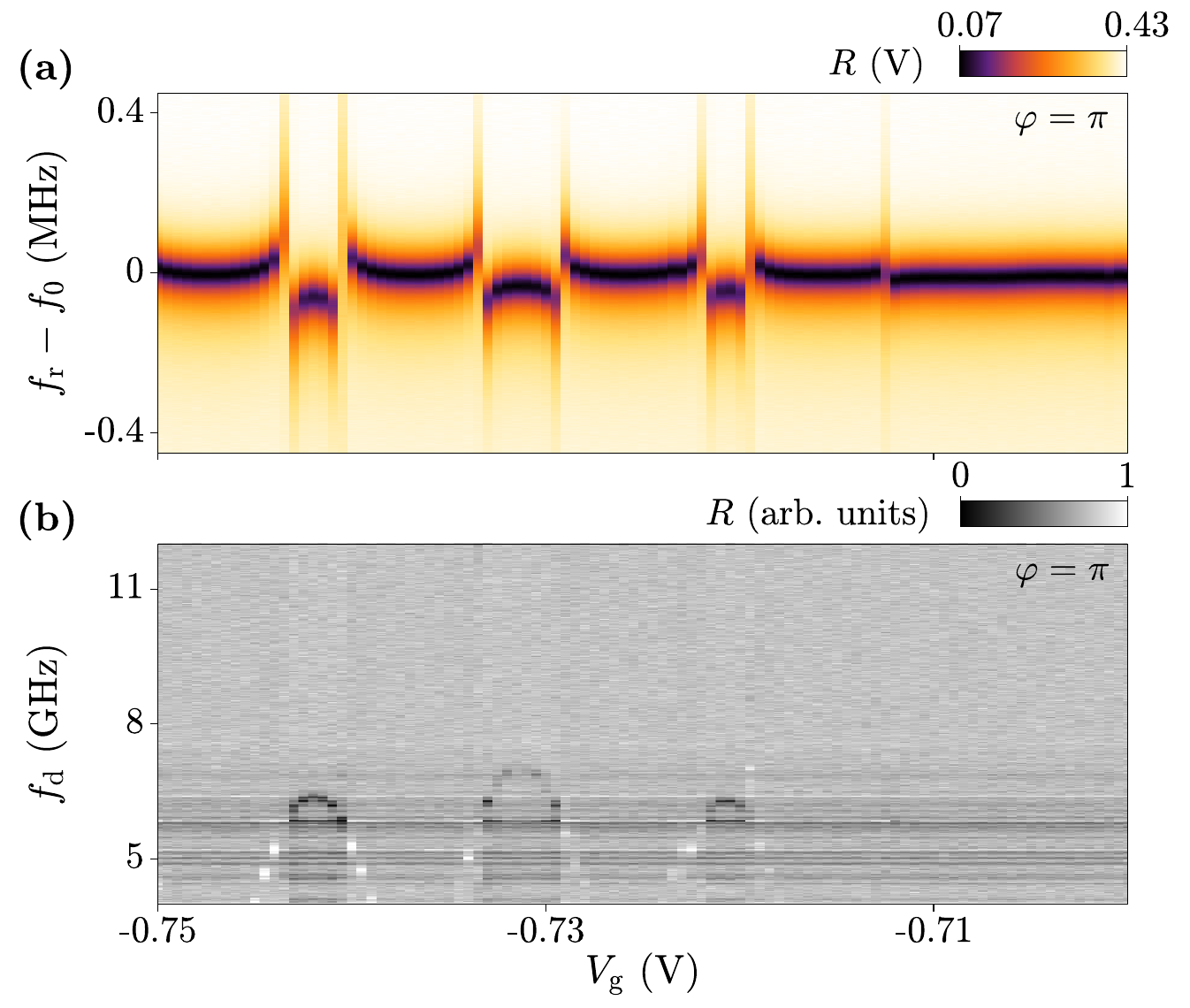}
	\caption{(a) Device~3: amplitude $R$ of the resonator transmission $S_{21}$ as a function of offset readout frequency $f_\mathrm{r}-f_0$ and gate voltage $V_\mathrm{g}$, measured at $\varphi=\pi$, with $f_0=5.83800$~GHz. (b) Amplitude $R$ as a function of drive frequency $f_\mathrm{d}$ ($P_\mathrm{d}=-16$~dBm) and gate voltage $V_\mathrm{g}$, measured together with (a).}
	\label{fig8}
\end{figure}

We also investigated a third device, with a junction length identical to device~1, $L=350$~nm, but with a slightly larger width, $W=150$~nm. This device had a bare resonance frequency $f_0=5.83800$~GHz. The internal and loaded quality factors were $Q_\mathrm{i}=189000$ and $Q_\mathrm{L}=36000$, respectively, at $V_\mathrm{g}=0$. The gate dependence of this device is depicted in Fig.~\ref{fig8}. Multiple anticrossings in Fig.~\ref{fig8}(a) indicate that the ABS transition frequency crossed the resonator several times, similar to the gate dependence of device~1 in Fig.~\ref{fig7}. The two-tone spectroscopy measurement in Fig.~\ref{fig8}(b) revealed an oscillating ABS, which crossed the resonator at gate voltages corresponding to anticrossings in Fig.~\ref{fig8}(a). The microwave line through which the drive tone was applied had 20~dB more attenuation than the microwave line that was used to apply the drive tone for device~1 and 2. Therefore, a higher drive power $P_\mathrm{d}$ was used for the measurement in Fig.~\ref{fig8}(b) than for the two-tone spectroscopy measurements in the main text. Moreover, for this device more standing waves were observed in the two-tone spectroscopy measurement. Nonetheless, the behavior of device~3 is similar to device~1.

\section*{APPENDIX F: THEORETICAL MODEL FOR FINITE JUNCTIONS}

To theoretically describe a finite-length JJ, we consider a few conducting channels, each of which can potentially host multiple ABSs. For the length scales relevant to our experiment, each channel can accommodate up to two ABSs. Furthermore, we observe transitions corresponding to ABSs in  two channels. This type of system has been studied in detail for InAs nanowires~\cite{Yeyati2017,Kurilovich2021,Tosi2019,Canadas2022}. In our Ge-based system, no spin splitting of the ABSs is observed, likely due to the weak spin-orbit interaction, consistent with strain-induced suppression of spin-orbit coupling in Ge samples with strong confinement in a single direction~\cite{Bosco2021}.

For the long junction, we take into account the effect of interactions, as QPs have a finite dwell time in the junction region. As predicted theoretically~\cite{Kurilovich2021} and observed experimentally~\cite{Fatemi2022,Canadas2022}, Coulomb interaction in intermediate-length junctions leads to finite shifts in the ABS spectrum. This interaction can be treated perturbatively due to the strong screening provided by the superconducting leads. The resulting effect is twofold. First, the wave function of a QP in the ABS of a ballistic junction extends into the superconducting leads, where the Coulomb interaction is screened. This reduces the effective interaction strength by a factor of $\Delta^2/\Gamma^2$, where $\Gamma$ is the coupling to the leads; for a ballistic channel, typically $\Gamma \gg \Delta$. Second, the QP charge is reduced, as it is a coherent superposition of an electron and a hole. This results in an energy-dependent effective charge. Consequently, the Coulomb energy is rescaled by $\sim[\Delta^2 - E_i^2(\varphi)]/\Delta^2$. Altogether, the Coulomb interaction for each ABS can be written as
\begin{equation}
	U_i(\varphi)=U_{c,i}\frac{\Delta^2-E_i^2(\varphi)}{\Gamma_i^2},
	\label{eq4}
\end{equation}
where $U_{c,i}=e^2/C_i$ is the charging energy determined by microscopic details of the corresponding channel. Thus, the coefficients $u_i$ in the main text are given by $u_i=(e^2/C_i)(1/\Gamma_i^2)$. Following Refs.~\cite{Kurilovich2021,Fatemi2022,Kurilovich2024}, we see that the lowest-order correction due to Coulomb interaction adds $U_i(\varphi)$ for every evenly-occupied ABS.

The phase dispersion of ABSs in a junction of intermediate length is more complex than for the short-junction regime \cite{Yeyati2017,Tosi2019}. However, the dispersion of the lowest-energy ABS in each channel ($E_1$ and $E_2$ in the main text) can be approximated by Eq.~\ref{eq1}. The higher-energy ABS in the experiment ($E_3$) resides in close proximity to the continuum, with weak phase dispersion. Therefore, it can be approximated using a weak cosine dispersion with maxima at $\pi(2n+1)$, resulting in
\begin{equation}
	E_{3}(\varphi)=\Delta - \widetilde{\Delta}_3[1+\cos(\varphi)].
	\label{eq5}
\end{equation}
The curvature sign of the phase dispersion indicates that this state is the second ABS in the channel. Due to its close proximity to the continuum, the effective charge of a QP in this ABS is very small. Hence, Coulomb interaction in this ABS can be neglected. 

Two different gate voltage regimes are presented in Fig.~\ref{fig4} of the main text. The ABS transition frequencies are obtained with the following parameters, where the induced superconducting gap $\Delta=14.5$~GHz:

\begin{center}
	\begin{tabular}{ |c|c|c|c|c|c|c|c| } 
		\hline
		Figure & $\tau_1$ & $\tau_2$ & $\widetilde{\Delta}_1/h$ & $\widetilde{\Delta}_2/h$& $\widetilde{\Delta}_3/h$& $u_1$ & $u_2$\\ 
		\hline
		\ref{fig4}(b) & 0.98 & 0.72  & 0.48 & 0.54 & 0.10 & 0.08 & 0.03\\ 
		\hline
		\ref{fig4}(d) & 0.62 & 0.51  & 0.50 & 0.52 & 0.07 & 0.01 & 0.01\\ 
		\hline
	\end{tabular}
\end{center}

\section*{APPENDIX G: DEVICE 2 DRIVE POWER DEPENDENCE}

\begin{figure}[H]
	\includegraphics[width=0.45\textwidth]{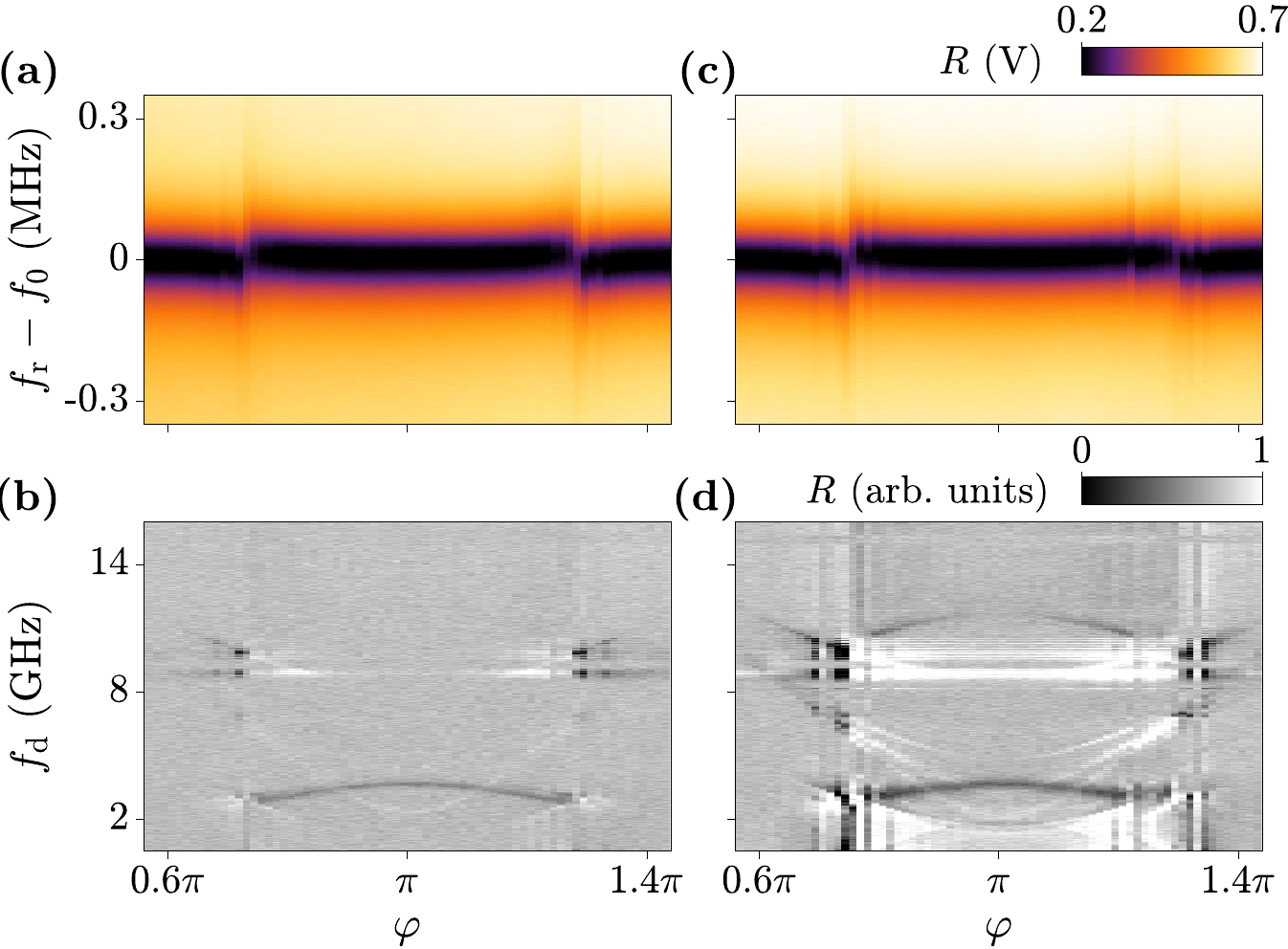}
	\caption{(a) Device~2: amplitude $R$ of the resonator transmission $S_{21}$ as a function of offset readout frequency $f_\mathrm{r}-f_0$ and phase $\varphi$, at the same gate voltage as Fig.~\ref{fig4}(a), $V_\mathrm{g}=-0.7990$~V, with $f_\mathrm{0}=6.86731$~GHz. (b) Amplitude $R$ of the resonator transmission $S_{21}$ as a function of drive frequency $f_\mathrm{d}$ ($P_\mathrm{d}=-35$~dBm) and phase $\varphi$, measured together with (a). Compared to Fig.~\ref{fig4}(a), fewer transitions are visible in the spectrum. (c,d) Same as (a,b), but with $P_\mathrm{d}=-25$~dBm. The same transitions as in Fig.~\ref{fig4}(a) were observed.}
	\label{fig10}
\end{figure}

Figure~\ref{fig10} depicts additional measurements of device~2, at the same gate voltage configuration as Fig.~\ref{fig4}(a), $V_\mathrm{g}=-0.7990$~V. The amplitude $R$ of the resonator transmission $S_{21}$ is shown as a function of the offset readout frequency $f_\mathrm{r}-f_0$ and the phase difference across the junction $\varphi$ in Fig.~\ref{fig10}(a). The resonator scans were acquired together with a two-tone spectroscopy measurement, shown in Fig.~\ref{fig10}(b), for which the drive power was set to $P_\mathrm{d}=-35$~dBm, lower than the drive power used for Fig.~\ref{fig4}(a), $P_\mathrm{d}=-28$~dBm. At this lower drive power, the PTs and the SQPT from $E_1$ to $E_2$ [Fig.~\ref{fig3}(d) and green dashed line in Fig.~\ref{fig4}(b)] are the most prominent transitions in the spectrum. This is expected for the PTs, with larger phase dispersion and therefore higher associated current, resulting in stronger coupling to the resonator in comparison to other transitions in the spectrum. Furthermore, the large linewidth and high intensity of the SQPT around $f_\mathrm{d}=4$~GHz in Fig.~\ref{fig4}(a) indicate that also this transition was associated to larger current and thus coupled strongly to the resonator. For the two-tone spectroscopy measurement shown in Fig.~\ref{fig10}(d), the drive power was increased to $P_\mathrm{d}=-25$~dBm. Despite applying higher drive power, the same transitions were observed in Fig.~\ref{fig10}(d) as in Fig.~\ref{fig4}(a). The anticrossings observed in Fig.~\ref{fig10}(c) deviate slightly from those in Fig.~\ref{fig10}(a), even though the drive power was off during these measurements. We suspect that this phenomenon is associated to instabilities in the gate voltage or applied flux. 

\section*{APPENDIX H: DEVICE 2 GATE DEPENDENCE}

\begin{figure}[H]
	\includegraphics[width=0.45\textwidth]{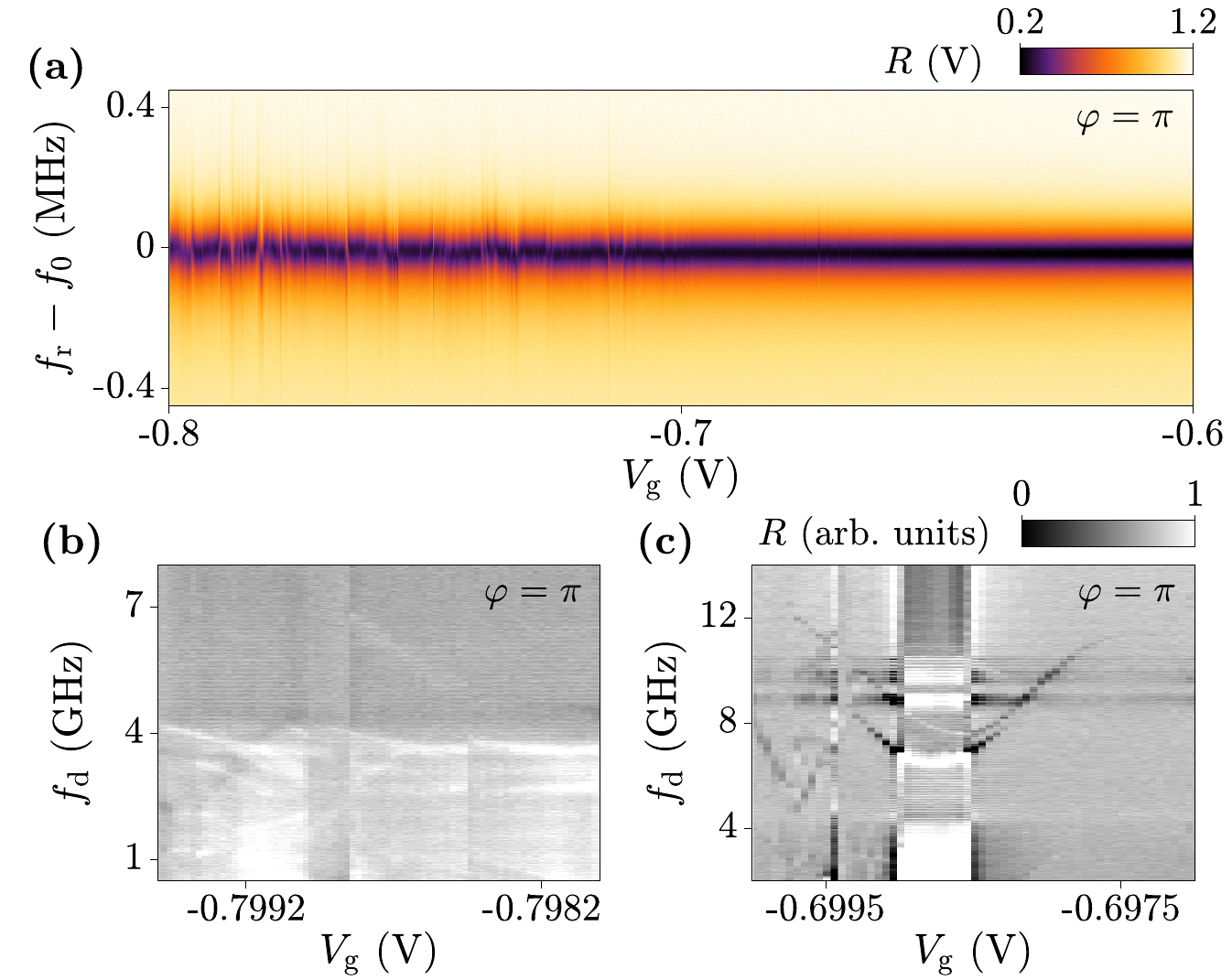}
	\caption{(a) Device 2: amplitude $R$ of the resonator transmission $S_{21}$ as a function of offset readout frequency $f_\mathrm{r}-f_0$ and gate voltage $V_\mathrm{g}$. (b) Amplitude $R$ of the resonator transmission $S_{21}$ as a function of drive frequency $f_\mathrm{d}$ ($P_\mathrm{d}=-35$~dBm) and gate voltage $V_\mathrm{g}$, measured with 250 averages. (c) Same as (b), but for a different gate voltage range and measured with $P_\mathrm{d}=-30$~dBm and 500 averages. All measurements were performed at $\varphi=\pi$.}
	\label{fig9}
\end{figure}

Figure~\ref{fig9}(a) depicts the gate response of device~2, measured over a wide range of gate voltages at phase $\varphi=\pi$. As the gate voltage $V_\mathrm{g}$ was swept to more negative values, the resonator trace became less pronounced, indicative of a decrease in the internal quality factor due to higher losses, which can be attributed to a larger current circulating in the superconducting loop of the device~\cite{Hinderling2023}. This larger current was paired with an increase in the number of ABSs in the junction, giving rise to more anticrossings in the left half of Fig.~\ref{fig9}(a). The combination of many ABSs and gate-voltage induced hysteresis caused gate voltage instabilities in this device. Therefore, the two-tone spectroscopy measurements in Fig.~\ref{fig4} cannot be associated to specific values of the gate voltages $V_\mathrm{g}$ and $V_\mathrm{s}$. Nonetheless, we present two gate dependent two-tone spectroscopy measurements in Fig.~\ref{fig9}(b) and (c), which were acquired at similar values of the gate voltage $V_\mathrm{g}$ as Figs.~\ref{fig4}(a) and (c), respectively, with $V_\mathrm{s}=0$. Many dispersing ABS transitions were observed in these gate voltage ranges, in agreement with the phase dependence measurements. Instabilities in the gate voltage are evident in Fig.~\ref{fig9}(b), where the spectrum seems to repeat itself below $V_\mathrm{g}=-0.7985$~V.

\bibliography{bibliography.bib}

\begin{thebibliography}{44}%
\makeatletter
\providecommand \@ifxundefined [1]{%
 \@ifx{#1\undefined}
}%
\providecommand \@ifnum [1]{%
 \ifnum #1\expandafter \@firstoftwo
 \else \expandafter \@secondoftwo
 \fi
}%
\providecommand \@ifx [1]{%
 \ifx #1\expandafter \@firstoftwo
 \else \expandafter \@secondoftwo
 \fi
}%
\providecommand \natexlab [1]{#1}%
\providecommand \enquote  [1]{``#1''}%
\providecommand \bibnamefont  [1]{#1}%
\providecommand \bibfnamefont [1]{#1}%
\providecommand \citenamefont [1]{#1}%
\providecommand \href@noop [0]{\@secondoftwo}%
\providecommand \href [0]{\begingroup \@sanitize@url \@href}%
\providecommand \@href[1]{\@@startlink{#1}\@@href}%
\providecommand \@@href[1]{\endgroup#1\@@endlink}%
\providecommand \@sanitize@url [0]{\catcode `\\12\catcode `\$12\catcode
  `\&12\catcode `\#12\catcode `\^12\catcode `\_12\catcode `\%12\relax}%
\providecommand \@@startlink[1]{}%
\providecommand \@@endlink[0]{}%
\providecommand \url  [0]{\begingroup\@sanitize@url \@url }%
\providecommand \@url [1]{\endgroup\@href {#1}{\urlprefix }}%
\providecommand \urlprefix  [0]{URL }%
\providecommand \Eprint [0]{\href }%
\providecommand \doibase [0]{https://doi.org/}%
\providecommand \selectlanguage [0]{\@gobble}%
\providecommand \bibinfo  [0]{\@secondoftwo}%
\providecommand \bibfield  [0]{\@secondoftwo}%
\providecommand \translation [1]{[#1]}%
\providecommand \BibitemOpen [0]{}%
\providecommand \bibitemStop [0]{}%
\providecommand \bibitemNoStop [0]{.\EOS\space}%
\providecommand \EOS [0]{\spacefactor3000\relax}%
\providecommand \BibitemShut  [1]{\csname bibitem#1\endcsname}%
\let\auto@bib@innerbib\@empty
%</preamble>
\bibitem [{\citenamefont {Andreev}(1964)}]{Andreev1964}%
  \BibitemOpen
  \bibfield  {author} {\bibinfo {author} {\bibfnamefont {A.~F.}\ \bibnamefont
  {Andreev}},\ }\bibfield  {title} {\bibinfo {title} {The thermal conductivity
  of the intermediate state in superconductors},\ }\href@noop {} {\bibfield
  {journal} {\bibinfo  {journal} {Journal of Experimental and Theoretical
  Physics}\ }\textbf {\bibinfo {volume} {46}},\ \bibinfo {pages} {1823}
  (\bibinfo {year} {1964})}\BibitemShut {NoStop}%
\bibitem [{\citenamefont {Blonder}\ \emph {et~al.}(1982)\citenamefont
  {Blonder}, \citenamefont {Tinkham},\ and\ \citenamefont
  {Klapwijk}}]{BTK1982}%
  \BibitemOpen
  \bibfield  {author} {\bibinfo {author} {\bibfnamefont {G.~E.}\ \bibnamefont
  {Blonder}}, \bibinfo {author} {\bibfnamefont {M.}~\bibnamefont {Tinkham}},\
  and\ \bibinfo {author} {\bibfnamefont {T.~M.}\ \bibnamefont {Klapwijk}},\
  }\bibfield  {title} {\bibinfo {title} {Transition from metallic to tunneling
  regimes in superconducting microconstrictions: {E}xcess current, charge
  imbalance, and supercurrent conversion},\ }\href
  {https://doi.org/10.1103/PhysRevB.25.4515} {\bibfield  {journal} {\bibinfo
  {journal} {Phys. Rev. B}\ }\textbf {\bibinfo {volume} {25}},\ \bibinfo
  {pages} {4515} (\bibinfo {year} {1982})}\BibitemShut {NoStop}%
\bibitem [{\citenamefont {Beenakker}(1991)}]{Beenakker1991_2}%
  \BibitemOpen
  \bibfield  {author} {\bibinfo {author} {\bibfnamefont {C.~W.~J.}\
  \bibnamefont {Beenakker}},\ }\bibfield  {title} {\bibinfo {title} {Universal
  limit of critical-current fluctuations in mesoscopic {J}osephson junctions},\
  }\href {https://doi.org/10.1103/PhysRevLett.67.3836} {\bibfield  {journal}
  {\bibinfo  {journal} {Phys. Rev. Lett.}\ }\textbf {\bibinfo {volume} {67}},\
  \bibinfo {pages} {3836} (\bibinfo {year} {1991})}\BibitemShut {NoStop}%
\bibitem [{\citenamefont {Furusaki}\ and\ \citenamefont
  {Tsukada}(1991)}]{Furusaki1991}%
  \BibitemOpen
  \bibfield  {author} {\bibinfo {author} {\bibfnamefont {A.}~\bibnamefont
  {Furusaki}}\ and\ \bibinfo {author} {\bibfnamefont {M.}~\bibnamefont
  {Tsukada}},\ }\bibfield  {title} {\bibinfo {title} {Current-carrying states
  in {J}osephson junctions},\ }\href
  {https://doi.org/10.1103/PhysRevB.43.10164} {\bibfield  {journal} {\bibinfo
  {journal} {Phys. Rev. B}\ }\textbf {\bibinfo {volume} {43}},\ \bibinfo
  {pages} {10164} (\bibinfo {year} {1991})}\BibitemShut {NoStop}%
\bibitem [{\citenamefont {Pillet}\ \emph {et~al.}(2013)\citenamefont {Pillet},
  \citenamefont {Joyez}, \citenamefont {\ifmmode~\check{Z}\else
  \v{Z}\fi{}itko},\ and\ \citenamefont {Goffman}}]{Pillet2013}%
  \BibitemOpen
  \bibfield  {author} {\bibinfo {author} {\bibfnamefont {J.-D.}\ \bibnamefont
  {Pillet}}, \bibinfo {author} {\bibfnamefont {P.}~\bibnamefont {Joyez}},
  \bibinfo {author} {\bibfnamefont {R.}~\bibnamefont {\ifmmode~\check{Z}\else
  \v{Z}\fi{}itko}},\ and\ \bibinfo {author} {\bibfnamefont {M.~F.}\
  \bibnamefont {Goffman}},\ }\bibfield  {title} {\bibinfo {title} {Tunneling
  spectroscopy of a single quantum dot coupled to a superconductor: From
  {K}ondo ridge to {A}ndreev bound states},\ }\href
  {https://doi.org/10.1103/PhysRevB.88.045101} {\bibfield  {journal} {\bibinfo
  {journal} {Phys. Rev. B}\ }\textbf {\bibinfo {volume} {88}},\ \bibinfo
  {pages} {045101} (\bibinfo {year} {2013})}\BibitemShut {NoStop}%
\bibitem [{\citenamefont {Nichele}\ \emph {et~al.}(2020)\citenamefont
  {Nichele}, \citenamefont {Portol\'es}, \citenamefont {Fornieri},
  \citenamefont {Whiticar}, \citenamefont {Drachmann}, \citenamefont {Gronin},
  \citenamefont {Wang}, \citenamefont {Gardner}, \citenamefont {Thomas},
  \citenamefont {Hatke}, \citenamefont {Manfra},\ and\ \citenamefont
  {Marcus}}]{Nichele2020}%
  \BibitemOpen
  \bibfield  {author} {\bibinfo {author} {\bibfnamefont {F.}~\bibnamefont
  {Nichele}}, \bibinfo {author} {\bibfnamefont {E.}~\bibnamefont {Portol\'es}},
  \bibinfo {author} {\bibfnamefont {A.}~\bibnamefont {Fornieri}}, \bibinfo
  {author} {\bibfnamefont {A.~M.}\ \bibnamefont {Whiticar}}, \bibinfo {author}
  {\bibfnamefont {A.~C.~C.}\ \bibnamefont {Drachmann}}, \bibinfo {author}
  {\bibfnamefont {S.}~\bibnamefont {Gronin}}, \bibinfo {author} {\bibfnamefont
  {T.}~\bibnamefont {Wang}}, \bibinfo {author} {\bibfnamefont {G.~C.}\
  \bibnamefont {Gardner}}, \bibinfo {author} {\bibfnamefont {C.}~\bibnamefont
  {Thomas}}, \bibinfo {author} {\bibfnamefont {A.~T.}\ \bibnamefont {Hatke}},
  \bibinfo {author} {\bibfnamefont {M.~J.}\ \bibnamefont {Manfra}},\ and\
  \bibinfo {author} {\bibfnamefont {C.~M.}\ \bibnamefont {Marcus}},\ }\bibfield
   {title} {\bibinfo {title} {Relating {A}ndreev bound states and supercurrents
  in hybrid {J}osephson junctions},\ }\href
  {https://doi.org/10.1103/PhysRevLett.124.226801} {\bibfield  {journal}
  {\bibinfo  {journal} {Phys. Rev. Lett.}\ }\textbf {\bibinfo {volume} {124}},\
  \bibinfo {pages} {226801} (\bibinfo {year} {2020})}\BibitemShut {NoStop}%
\bibitem [{\citenamefont {Coraiola}\ \emph {et~al.}(2023)\citenamefont
  {Coraiola}, \citenamefont {Haxell}, \citenamefont {Sabonis}, \citenamefont
  {Weisbrich}, \citenamefont {Svetogorov}, \citenamefont {Hinderling},
  \citenamefont {ten Kate}, \citenamefont {Cheah}, \citenamefont {Krizek},
  \citenamefont {Schott}, \citenamefont {Wegscheider}, \citenamefont {Cuevas},
  \citenamefont {Belzig},\ and\ \citenamefont {Nichele}}]{Coraiola2023}%
  \BibitemOpen
  \bibfield  {author} {\bibinfo {author} {\bibfnamefont {M.}~\bibnamefont
  {Coraiola}}, \bibinfo {author} {\bibfnamefont {D.~Z.}\ \bibnamefont
  {Haxell}}, \bibinfo {author} {\bibfnamefont {D.}~\bibnamefont {Sabonis}},
  \bibinfo {author} {\bibfnamefont {H.}~\bibnamefont {Weisbrich}}, \bibinfo
  {author} {\bibfnamefont {A.~E.}\ \bibnamefont {Svetogorov}}, \bibinfo
  {author} {\bibfnamefont {M.}~\bibnamefont {Hinderling}}, \bibinfo {author}
  {\bibfnamefont {S.~C.}\ \bibnamefont {ten Kate}}, \bibinfo {author}
  {\bibfnamefont {E.}~\bibnamefont {Cheah}}, \bibinfo {author} {\bibfnamefont
  {F.}~\bibnamefont {Krizek}}, \bibinfo {author} {\bibfnamefont
  {R.}~\bibnamefont {Schott}}, \bibinfo {author} {\bibfnamefont
  {W.}~\bibnamefont {Wegscheider}}, \bibinfo {author} {\bibfnamefont {J.~C.}\
  \bibnamefont {Cuevas}}, \bibinfo {author} {\bibfnamefont {W.}~\bibnamefont
  {Belzig}},\ and\ \bibinfo {author} {\bibfnamefont {F.}~\bibnamefont
  {Nichele}},\ }\bibfield  {title} {\bibinfo {title} {Phase-engineering the
  {A}ndreev band structure of a three-terminal {J}osephson junction},\ }\href
  {https://doi.org/10.1038/s41467-023-42356-6} {\bibfield  {journal} {\bibinfo
  {journal} {Nat. Commun.}\ }\textbf {\bibinfo {volume} {14}},\ \bibinfo
  {pages} {6784} (\bibinfo {year} {2023})}\BibitemShut {NoStop}%
\bibitem [{\citenamefont {Janvier}\ \emph {et~al.}(2015)\citenamefont
  {Janvier}, \citenamefont {Tosi}, \citenamefont {Bretheau}, \citenamefont
  {\c{C}. {\"O}.~Girit}, \citenamefont {Stern}, \citenamefont {Bertet},
  \citenamefont {Joyez}, \citenamefont {Vion}, \citenamefont {Esteve},
  \citenamefont {Goffman}, \citenamefont {Pothier},\ and\ \citenamefont
  {Urbina}}]{Janvier2015}%
  \BibitemOpen
  \bibfield  {author} {\bibinfo {author} {\bibfnamefont {C.}~\bibnamefont
  {Janvier}}, \bibinfo {author} {\bibfnamefont {L.}~\bibnamefont {Tosi}},
  \bibinfo {author} {\bibfnamefont {L.}~\bibnamefont {Bretheau}}, \bibinfo
  {author} {\bibnamefont {\c{C}. {\"O}.~Girit}}, \bibinfo {author}
  {\bibfnamefont {M.}~\bibnamefont {Stern}}, \bibinfo {author} {\bibfnamefont
  {P.}~\bibnamefont {Bertet}}, \bibinfo {author} {\bibfnamefont
  {P.}~\bibnamefont {Joyez}}, \bibinfo {author} {\bibfnamefont
  {D.}~\bibnamefont {Vion}}, \bibinfo {author} {\bibfnamefont {D.}~\bibnamefont
  {Esteve}}, \bibinfo {author} {\bibfnamefont {M.~F.}\ \bibnamefont {Goffman}},
  \bibinfo {author} {\bibfnamefont {H.}~\bibnamefont {Pothier}},\ and\ \bibinfo
  {author} {\bibfnamefont {C.}~\bibnamefont {Urbina}},\ }\bibfield  {title}
  {\bibinfo {title} {Coherent manipulation of {A}ndreev states in
  superconducting atomic contacts},\ }\href
  {https://doi.org/10.1126/science.aab2179} {\bibfield  {journal} {\bibinfo
  {journal} {Science}\ }\textbf {\bibinfo {volume} {349}},\ \bibinfo {pages}
  {1199} (\bibinfo {year} {2015})}\BibitemShut {NoStop}%
\bibitem [{\citenamefont {Hays}\ \emph {et~al.}(2018)\citenamefont {Hays},
  \citenamefont {de~Lange}, \citenamefont {Serniak}, \citenamefont {van
  Woerkom}, \citenamefont {Bouman}, \citenamefont {Krogstrup}, \citenamefont
  {Nyg\aa{}rd}, \citenamefont {Geresdi},\ and\ \citenamefont
  {Devoret}}]{Hays2018}%
  \BibitemOpen
  \bibfield  {author} {\bibinfo {author} {\bibfnamefont {M.}~\bibnamefont
  {Hays}}, \bibinfo {author} {\bibfnamefont {G.}~\bibnamefont {de~Lange}},
  \bibinfo {author} {\bibfnamefont {K.}~\bibnamefont {Serniak}}, \bibinfo
  {author} {\bibfnamefont {D.~J.}\ \bibnamefont {van Woerkom}}, \bibinfo
  {author} {\bibfnamefont {D.}~\bibnamefont {Bouman}}, \bibinfo {author}
  {\bibfnamefont {P.}~\bibnamefont {Krogstrup}}, \bibinfo {author}
  {\bibfnamefont {J.}~\bibnamefont {Nyg\aa{}rd}}, \bibinfo {author}
  {\bibfnamefont {A.}~\bibnamefont {Geresdi}},\ and\ \bibinfo {author}
  {\bibfnamefont {M.~H.}\ \bibnamefont {Devoret}},\ }\bibfield  {title}
  {\bibinfo {title} {Direct microwave measurement of {A}ndreev-bound-state
  dynamics in a semiconductor-nanowire {J}osephson junction},\ }\href
  {https://doi.org/10.1103/PhysRevLett.121.047001} {\bibfield  {journal}
  {\bibinfo  {journal} {Phys. Rev. Lett.}\ }\textbf {\bibinfo {volume} {121}},\
  \bibinfo {pages} {047001} (\bibinfo {year} {2018})}\BibitemShut {NoStop}%
\bibitem [{\citenamefont {Tosi}\ \emph {et~al.}(2019)\citenamefont {Tosi},
  \citenamefont {Metzger}, \citenamefont {Goffman}, \citenamefont {Urbina},
  \citenamefont {Pothier}, \citenamefont {Park}, \citenamefont {Yeyati},
  \citenamefont {Nyg\aa{}rd},\ and\ \citenamefont {Krogstrup}}]{Tosi2019}%
  \BibitemOpen
  \bibfield  {author} {\bibinfo {author} {\bibfnamefont {L.}~\bibnamefont
  {Tosi}}, \bibinfo {author} {\bibfnamefont {C.}~\bibnamefont {Metzger}},
  \bibinfo {author} {\bibfnamefont {M.~F.}\ \bibnamefont {Goffman}}, \bibinfo
  {author} {\bibfnamefont {C.}~\bibnamefont {Urbina}}, \bibinfo {author}
  {\bibfnamefont {H.}~\bibnamefont {Pothier}}, \bibinfo {author} {\bibfnamefont
  {S.}~\bibnamefont {Park}}, \bibinfo {author} {\bibfnamefont {A.~L.}\
  \bibnamefont {Yeyati}}, \bibinfo {author} {\bibfnamefont {J.}~\bibnamefont
  {Nyg\aa{}rd}},\ and\ \bibinfo {author} {\bibfnamefont {P.}~\bibnamefont
  {Krogstrup}},\ }\bibfield  {title} {\bibinfo {title} {Spin-orbit splitting of
  {A}ndreev states revealed by microwave spectroscopy},\ }\href
  {https://doi.org/10.1103/PhysRevX.9.011010} {\bibfield  {journal} {\bibinfo
  {journal} {Phys. Rev. X}\ }\textbf {\bibinfo {volume} {9}},\ \bibinfo {pages}
  {011010} (\bibinfo {year} {2019})}\BibitemShut {NoStop}%
\bibitem [{\citenamefont {Hays}\ \emph {et~al.}(2020)\citenamefont {Hays},
  \citenamefont {Fatemi}, \citenamefont {Serniak}, \citenamefont {Bouman},
  \citenamefont {Diamond}, \citenamefont {de~Lange}, \citenamefont {Krogstrup},
  \citenamefont {Nyg{\aa}rd}, \citenamefont {Geresdi},\ and\ \citenamefont
  {Devoret}}]{Hays2020}%
  \BibitemOpen
  \bibfield  {author} {\bibinfo {author} {\bibfnamefont {M.}~\bibnamefont
  {Hays}}, \bibinfo {author} {\bibfnamefont {V.}~\bibnamefont {Fatemi}},
  \bibinfo {author} {\bibfnamefont {K.}~\bibnamefont {Serniak}}, \bibinfo
  {author} {\bibfnamefont {D.}~\bibnamefont {Bouman}}, \bibinfo {author}
  {\bibfnamefont {S.}~\bibnamefont {Diamond}}, \bibinfo {author} {\bibfnamefont
  {G.}~\bibnamefont {de~Lange}}, \bibinfo {author} {\bibfnamefont
  {P.}~\bibnamefont {Krogstrup}}, \bibinfo {author} {\bibfnamefont
  {J.}~\bibnamefont {Nyg{\aa}rd}}, \bibinfo {author} {\bibfnamefont
  {A.}~\bibnamefont {Geresdi}},\ and\ \bibinfo {author} {\bibfnamefont {M.~H.}\
  \bibnamefont {Devoret}},\ }\bibfield  {title} {\bibinfo {title} {Continuous
  monitoring of a trapped superconducting spin},\ }\href
  {https://doi.org/10.1038/s41567-020-0952-3} {\bibfield  {journal} {\bibinfo
  {journal} {Nat. Phys.}\ }\textbf {\bibinfo {volume} {16}},\ \bibinfo {pages}
  {1103} (\bibinfo {year} {2020})}\BibitemShut {NoStop}%
\bibitem [{\citenamefont {Hays}\ \emph {et~al.}(2021)\citenamefont {Hays},
  \citenamefont {Fatemi}, \citenamefont {Bouman}, \citenamefont {Cerrillo},
  \citenamefont {Diamond}, \citenamefont {Serniak}, \citenamefont {Connolly},
  \citenamefont {Krogstrup}, \citenamefont {Nygaard}, \citenamefont
  {Levy~Yeyati} \emph {et~al.}}]{Hays2021}%
  \BibitemOpen
  \bibfield  {author} {\bibinfo {author} {\bibfnamefont {M.}~\bibnamefont
  {Hays}}, \bibinfo {author} {\bibfnamefont {V.}~\bibnamefont {Fatemi}},
  \bibinfo {author} {\bibfnamefont {D.}~\bibnamefont {Bouman}}, \bibinfo
  {author} {\bibfnamefont {J.}~\bibnamefont {Cerrillo}}, \bibinfo {author}
  {\bibfnamefont {S.}~\bibnamefont {Diamond}}, \bibinfo {author} {\bibfnamefont
  {K.}~\bibnamefont {Serniak}}, \bibinfo {author} {\bibfnamefont
  {T.}~\bibnamefont {Connolly}}, \bibinfo {author} {\bibfnamefont
  {P.}~\bibnamefont {Krogstrup}}, \bibinfo {author} {\bibfnamefont
  {J.}~\bibnamefont {Nygaard}}, \bibinfo {author} {\bibfnamefont
  {A.}~\bibnamefont {Levy~Yeyati}}, \emph {et~al.},\ }\bibfield  {title}
  {\bibinfo {title} {Coherent manipulation of an {A}ndreev spin qubit},\
  }\href@noop {} {\bibfield  {journal} {\bibinfo  {journal} {Science}\ }\textbf
  {\bibinfo {volume} {373}},\ \bibinfo {pages} {430} (\bibinfo {year}
  {2021})}\BibitemShut {NoStop}%
\bibitem [{\citenamefont {Chidambaram}\ \emph {et~al.}(2022)\citenamefont
  {Chidambaram}, \citenamefont {Kringh\o{}j}, \citenamefont {Casparis},
  \citenamefont {Kuemmeth}, \citenamefont {Wang}, \citenamefont {Thomas},
  \citenamefont {Gronin}, \citenamefont {Gardner}, \citenamefont {Cui},
  \citenamefont {Liu}, \citenamefont {Moors}, \citenamefont {Manfra},
  \citenamefont {Petersson},\ and\ \citenamefont {Connolly}}]{Chidambaram2022}%
  \BibitemOpen
  \bibfield  {author} {\bibinfo {author} {\bibfnamefont {V.}~\bibnamefont
  {Chidambaram}}, \bibinfo {author} {\bibfnamefont {A.}~\bibnamefont
  {Kringh\o{}j}}, \bibinfo {author} {\bibfnamefont {L.}~\bibnamefont
  {Casparis}}, \bibinfo {author} {\bibfnamefont {F.}~\bibnamefont {Kuemmeth}},
  \bibinfo {author} {\bibfnamefont {T.}~\bibnamefont {Wang}}, \bibinfo {author}
  {\bibfnamefont {C.}~\bibnamefont {Thomas}}, \bibinfo {author} {\bibfnamefont
  {S.}~\bibnamefont {Gronin}}, \bibinfo {author} {\bibfnamefont {G.~C.}\
  \bibnamefont {Gardner}}, \bibinfo {author} {\bibfnamefont {Z.}~\bibnamefont
  {Cui}}, \bibinfo {author} {\bibfnamefont {C.}~\bibnamefont {Liu}}, \bibinfo
  {author} {\bibfnamefont {K.}~\bibnamefont {Moors}}, \bibinfo {author}
  {\bibfnamefont {M.~J.}\ \bibnamefont {Manfra}}, \bibinfo {author}
  {\bibfnamefont {K.~D.}\ \bibnamefont {Petersson}},\ and\ \bibinfo {author}
  {\bibfnamefont {M.~R.}\ \bibnamefont {Connolly}},\ }\bibfield  {title}
  {\bibinfo {title} {Microwave sensing of {A}ndreev bound states in a
  gate-defined superconducting quantum point contact},\ }\href
  {https://doi.org/10.1103/PhysRevResearch.4.023170} {\bibfield  {journal}
  {\bibinfo  {journal} {Phys. Rev. Res.}\ }\textbf {\bibinfo {volume} {4}},\
  \bibinfo {pages} {023170} (\bibinfo {year} {2022})}\BibitemShut {NoStop}%
\bibitem [{\citenamefont {Matute-Ca\~nadas}\ \emph {et~al.}(2022)\citenamefont
  {Matute-Ca\~nadas}, \citenamefont {Metzger}, \citenamefont {Park},
  \citenamefont {Tosi}, \citenamefont {Krogstrup}, \citenamefont {Nyg\aa{}rd},
  \citenamefont {Goffman}, \citenamefont {Urbina}, \citenamefont {Pothier},\
  and\ \citenamefont {Yeyati}}]{Canadas2022}%
  \BibitemOpen
  \bibfield  {author} {\bibinfo {author} {\bibfnamefont {F.~J.}\ \bibnamefont
  {Matute-Ca\~nadas}}, \bibinfo {author} {\bibfnamefont {C.}~\bibnamefont
  {Metzger}}, \bibinfo {author} {\bibfnamefont {S.}~\bibnamefont {Park}},
  \bibinfo {author} {\bibfnamefont {L.}~\bibnamefont {Tosi}}, \bibinfo {author}
  {\bibfnamefont {P.}~\bibnamefont {Krogstrup}}, \bibinfo {author}
  {\bibfnamefont {J.}~\bibnamefont {Nyg\aa{}rd}}, \bibinfo {author}
  {\bibfnamefont {M.~F.}\ \bibnamefont {Goffman}}, \bibinfo {author}
  {\bibfnamefont {C.}~\bibnamefont {Urbina}}, \bibinfo {author} {\bibfnamefont
  {H.}~\bibnamefont {Pothier}},\ and\ \bibinfo {author} {\bibfnamefont {A.~L.}\
  \bibnamefont {Yeyati}},\ }\bibfield  {title} {\bibinfo {title} {Signatures of
  interactions in the {A}ndreev spectrum of nanowire {J}osephson junctions},\
  }\href {https://doi.org/10.1103/PhysRevLett.128.197702} {\bibfield  {journal}
  {\bibinfo  {journal} {Phys. Rev. Lett.}\ }\textbf {\bibinfo {volume} {128}},\
  \bibinfo {pages} {197702} (\bibinfo {year} {2022})}\BibitemShut {NoStop}%
\bibitem [{\citenamefont {Zellekens}\ \emph {et~al.}(2022)\citenamefont
  {Zellekens}, \citenamefont {Deacon}, \citenamefont {Perla}, \citenamefont
  {Gr{\"u}tzmacher}, \citenamefont {Lepsa}, \citenamefont {Sch{\"a}pers},\ and\
  \citenamefont {Ishibashi}}]{Zellekens2022}%
  \BibitemOpen
  \bibfield  {author} {\bibinfo {author} {\bibfnamefont {P.}~\bibnamefont
  {Zellekens}}, \bibinfo {author} {\bibfnamefont {R.~S.}\ \bibnamefont
  {Deacon}}, \bibinfo {author} {\bibfnamefont {P.}~\bibnamefont {Perla}},
  \bibinfo {author} {\bibfnamefont {D.}~\bibnamefont {Gr{\"u}tzmacher}},
  \bibinfo {author} {\bibfnamefont {M.~I.}\ \bibnamefont {Lepsa}}, \bibinfo
  {author} {\bibfnamefont {T.}~\bibnamefont {Sch{\"a}pers}},\ and\ \bibinfo
  {author} {\bibfnamefont {K.}~\bibnamefont {Ishibashi}},\ }\bibfield  {title}
  {\bibinfo {title} {Microwave spectroscopy of {A}ndreev states in {I}n{A}s
  nanowire-based hybrid junctions using a flip-chip layout},\ }\href
  {https://doi.org/10.1038/s42005-022-01035-6} {\bibfield  {journal} {\bibinfo
  {journal} {Commun. Phys.}\ }\textbf {\bibinfo {volume} {5}},\ \bibinfo
  {pages} {267} (\bibinfo {year} {2022})}\BibitemShut {NoStop}%
\bibitem [{\citenamefont {Fatemi}\ \emph {et~al.}(2022)\citenamefont {Fatemi},
  \citenamefont {Kurilovich}, \citenamefont {Hays}, \citenamefont {Bouman},
  \citenamefont {Connolly}, \citenamefont {Diamond}, \citenamefont {Frattini},
  \citenamefont {Kurilovich}, \citenamefont {Krogstrup}, \citenamefont
  {Nyg\aa{}rd}, \citenamefont {Geresdi}, \citenamefont {Glazman},\ and\
  \citenamefont {Devoret}}]{Fatemi2022}%
  \BibitemOpen
  \bibfield  {author} {\bibinfo {author} {\bibfnamefont {V.}~\bibnamefont
  {Fatemi}}, \bibinfo {author} {\bibfnamefont {P.~D.}\ \bibnamefont
  {Kurilovich}}, \bibinfo {author} {\bibfnamefont {M.}~\bibnamefont {Hays}},
  \bibinfo {author} {\bibfnamefont {D.}~\bibnamefont {Bouman}}, \bibinfo
  {author} {\bibfnamefont {T.}~\bibnamefont {Connolly}}, \bibinfo {author}
  {\bibfnamefont {S.}~\bibnamefont {Diamond}}, \bibinfo {author} {\bibfnamefont
  {N.~E.}\ \bibnamefont {Frattini}}, \bibinfo {author} {\bibfnamefont {V.~D.}\
  \bibnamefont {Kurilovich}}, \bibinfo {author} {\bibfnamefont
  {P.}~\bibnamefont {Krogstrup}}, \bibinfo {author} {\bibfnamefont
  {J.}~\bibnamefont {Nyg\aa{}rd}}, \bibinfo {author} {\bibfnamefont
  {A.}~\bibnamefont {Geresdi}}, \bibinfo {author} {\bibfnamefont {L.~I.}\
  \bibnamefont {Glazman}},\ and\ \bibinfo {author} {\bibfnamefont {M.~H.}\
  \bibnamefont {Devoret}},\ }\bibfield  {title} {\bibinfo {title} {Microwave
  susceptibility observation of interacting many-body {A}ndreev states},\
  }\href {https://doi.org/10.1103/PhysRevLett.129.227701} {\bibfield  {journal}
  {\bibinfo  {journal} {Phys. Rev. Lett.}\ }\textbf {\bibinfo {volume} {129}},\
  \bibinfo {pages} {227701} (\bibinfo {year} {2022})}\BibitemShut {NoStop}%
\bibitem [{\citenamefont {Hinderling}\ \emph {et~al.}(2023)\citenamefont
  {Hinderling}, \citenamefont {Sabonis}, \citenamefont {Paredes}, \citenamefont
  {Haxell}, \citenamefont {Coraiola}, \citenamefont {ten Kate}, \citenamefont
  {Cheah}, \citenamefont {Krizek}, \citenamefont {Schott}, \citenamefont
  {Wegscheider},\ and\ \citenamefont {Nichele}}]{Hinderling2023}%
  \BibitemOpen
  \bibfield  {author} {\bibinfo {author} {\bibfnamefont {M.}~\bibnamefont
  {Hinderling}}, \bibinfo {author} {\bibfnamefont {D.}~\bibnamefont {Sabonis}},
  \bibinfo {author} {\bibfnamefont {S.}~\bibnamefont {Paredes}}, \bibinfo
  {author} {\bibfnamefont {D.}~\bibnamefont {Haxell}}, \bibinfo {author}
  {\bibfnamefont {M.}~\bibnamefont {Coraiola}}, \bibinfo {author}
  {\bibfnamefont {S.}~\bibnamefont {ten Kate}}, \bibinfo {author}
  {\bibfnamefont {E.}~\bibnamefont {Cheah}}, \bibinfo {author} {\bibfnamefont
  {F.}~\bibnamefont {Krizek}}, \bibinfo {author} {\bibfnamefont
  {R.}~\bibnamefont {Schott}}, \bibinfo {author} {\bibfnamefont
  {W.}~\bibnamefont {Wegscheider}},\ and\ \bibinfo {author} {\bibfnamefont
  {F.}~\bibnamefont {Nichele}},\ }\bibfield  {title} {\bibinfo {title}
  {Flip-chip-based microwave spectroscopy of {A}ndreev bound states in a planar
  {J}osephson junction},\ }\href
  {https://doi.org/10.1103/PhysRevApplied.19.054026} {\bibfield  {journal}
  {\bibinfo  {journal} {Phys. Rev. Appl.}\ }\textbf {\bibinfo {volume} {19}},\
  \bibinfo {pages} {054026} (\bibinfo {year} {2023})}\BibitemShut {NoStop}%
\bibitem [{\citenamefont {Wesdorp}\ \emph {et~al.}(2024)\citenamefont
  {Wesdorp}, \citenamefont {Matute-Ca\~nadas}, \citenamefont {Vaartjes},
  \citenamefont {Gr\"unhaupt}, \citenamefont {Laeven}, \citenamefont {Roelofs},
  \citenamefont {Splitthoff}, \citenamefont {Pita-Vidal}, \citenamefont
  {Bargerbos}, \citenamefont {van Woerkom}, \citenamefont {Krogstrup},
  \citenamefont {Kouwenhoven}, \citenamefont {Andersen}, \citenamefont
  {Yeyati}, \citenamefont {van Heck},\ and\ \citenamefont
  {de~Lange}}]{Wesdorp2024}%
  \BibitemOpen
  \bibfield  {author} {\bibinfo {author} {\bibfnamefont {J.~J.}\ \bibnamefont
  {Wesdorp}}, \bibinfo {author} {\bibfnamefont {F.~J.}\ \bibnamefont
  {Matute-Ca\~nadas}}, \bibinfo {author} {\bibfnamefont {A.}~\bibnamefont
  {Vaartjes}}, \bibinfo {author} {\bibfnamefont {L.}~\bibnamefont
  {Gr\"unhaupt}}, \bibinfo {author} {\bibfnamefont {T.}~\bibnamefont {Laeven}},
  \bibinfo {author} {\bibfnamefont {S.}~\bibnamefont {Roelofs}}, \bibinfo
  {author} {\bibfnamefont {L.~J.}\ \bibnamefont {Splitthoff}}, \bibinfo
  {author} {\bibfnamefont {M.}~\bibnamefont {Pita-Vidal}}, \bibinfo {author}
  {\bibfnamefont {A.}~\bibnamefont {Bargerbos}}, \bibinfo {author}
  {\bibfnamefont {D.~J.}\ \bibnamefont {van Woerkom}}, \bibinfo {author}
  {\bibfnamefont {P.}~\bibnamefont {Krogstrup}}, \bibinfo {author}
  {\bibfnamefont {L.~P.}\ \bibnamefont {Kouwenhoven}}, \bibinfo {author}
  {\bibfnamefont {C.~K.}\ \bibnamefont {Andersen}}, \bibinfo {author}
  {\bibfnamefont {A.~L.}\ \bibnamefont {Yeyati}}, \bibinfo {author}
  {\bibfnamefont {B.}~\bibnamefont {van Heck}},\ and\ \bibinfo {author}
  {\bibfnamefont {G.}~\bibnamefont {de~Lange}},\ }\bibfield  {title} {\bibinfo
  {title} {Microwave spectroscopy of interacting {A}ndreev spins},\ }\href
  {https://doi.org/10.1103/PhysRevB.109.045302} {\bibfield  {journal} {\bibinfo
   {journal} {Phys. Rev. B}\ }\textbf {\bibinfo {volume} {109}},\ \bibinfo
  {pages} {045302} (\bibinfo {year} {2024})}\BibitemShut {NoStop}%
\bibitem [{\citenamefont {Hinderling}\ \emph
  {et~al.}(2024{\natexlab{a}})\citenamefont {Hinderling}, \citenamefont {ten
  Kate}, \citenamefont {Coraiola}, \citenamefont {Haxell}, \citenamefont
  {Stiefel}, \citenamefont {Mergenthaler}, \citenamefont {Paredes},
  \citenamefont {Bedell}, \citenamefont {Sabonis},\ and\ \citenamefont
  {Nichele}}]{Hinderling2024Ge}%
  \BibitemOpen
  \bibfield  {author} {\bibinfo {author} {\bibfnamefont {M.}~\bibnamefont
  {Hinderling}}, \bibinfo {author} {\bibfnamefont {S.~C.}\ \bibnamefont {ten
  Kate}}, \bibinfo {author} {\bibfnamefont {M.}~\bibnamefont {Coraiola}},
  \bibinfo {author} {\bibfnamefont {D.}~\bibnamefont {Haxell}}, \bibinfo
  {author} {\bibfnamefont {M.}~\bibnamefont {Stiefel}}, \bibinfo {author}
  {\bibfnamefont {M.}~\bibnamefont {Mergenthaler}}, \bibinfo {author}
  {\bibfnamefont {S.}~\bibnamefont {Paredes}}, \bibinfo {author} {\bibfnamefont
  {S.}~\bibnamefont {Bedell}}, \bibinfo {author} {\bibfnamefont
  {D.}~\bibnamefont {Sabonis}},\ and\ \bibinfo {author} {\bibfnamefont
  {F.}~\bibnamefont {Nichele}},\ }\bibfield  {title} {\bibinfo {title} {Direct
  microwave spectroscopy of {A}ndreev bound states in planar $\mathrm{Ge}$
  {J}osephson junctions},\ }\href {https://doi.org/10.1103/PRXQuantum.5.030357}
  {\bibfield  {journal} {\bibinfo  {journal} {PRX Quantum}\ }\textbf {\bibinfo
  {volume} {5}},\ \bibinfo {pages} {030357} (\bibinfo {year}
  {2024}{\natexlab{a}})}\BibitemShut {NoStop}%
\bibitem [{\citenamefont {Elfeky}\ \emph {et~al.}(2025)\citenamefont {Elfeky},
  \citenamefont {Dindial}, \citenamefont {Brand\~ao}, \citenamefont {Pekerten},
  \citenamefont {Lee}, \citenamefont {Strickland}, \citenamefont {Strohbeen},
  \citenamefont {Danilenko}, \citenamefont {Baker}, \citenamefont {Mikalsen},
  \citenamefont {Schiela}, \citenamefont {Liang}, \citenamefont {Issokson},
  \citenamefont {Levy}, \citenamefont {\ifmmode \check{Z}\else
  \v{Z}\fi{}uti\ifmmode~\acute{c}\else \'{c}\fi{}},\ and\ \citenamefont
  {Shabani}}]{Elfeky2025}%
  \BibitemOpen
  \bibfield  {author} {\bibinfo {author} {\bibfnamefont {B.~H.}\ \bibnamefont
  {Elfeky}}, \bibinfo {author} {\bibfnamefont {K.}~\bibnamefont {Dindial}},
  \bibinfo {author} {\bibfnamefont {D.~S.}\ \bibnamefont {Brand\~ao}}, \bibinfo
  {author} {\bibfnamefont {B.~m.~c.}\ \bibnamefont {Pekerten}}, \bibinfo
  {author} {\bibfnamefont {J.}~\bibnamefont {Lee}}, \bibinfo {author}
  {\bibfnamefont {W.~M.}\ \bibnamefont {Strickland}}, \bibinfo {author}
  {\bibfnamefont {P.~J.}\ \bibnamefont {Strohbeen}}, \bibinfo {author}
  {\bibfnamefont {A.}~\bibnamefont {Danilenko}}, \bibinfo {author}
  {\bibfnamefont {L.}~\bibnamefont {Baker}}, \bibinfo {author} {\bibfnamefont
  {M.}~\bibnamefont {Mikalsen}}, \bibinfo {author} {\bibfnamefont
  {W.}~\bibnamefont {Schiela}}, \bibinfo {author} {\bibfnamefont
  {Z.}~\bibnamefont {Liang}}, \bibinfo {author} {\bibfnamefont
  {J.}~\bibnamefont {Issokson}}, \bibinfo {author} {\bibfnamefont
  {I.}~\bibnamefont {Levy}}, \bibinfo {author} {\bibfnamefont {I.}~\bibnamefont
  {\ifmmode \check{Z}\else \v{Z}\fi{}uti\ifmmode~\acute{c}\else \'{c}\fi{}}},\
  and\ \bibinfo {author} {\bibfnamefont {J.}~\bibnamefont {Shabani}},\
  }\bibfield  {title} {\bibinfo {title} {Microwave {A}ndreev bound state
  spectroscopy in a semiconductor-based planar {J}osephson junction},\ }\href
  {https://doi.org/10.1103/PhysRevResearch.7.013248} {\bibfield  {journal}
  {\bibinfo  {journal} {Phys. Rev. Res.}\ }\textbf {\bibinfo {volume} {7}},\
  \bibinfo {pages} {013248} (\bibinfo {year} {2025})}\BibitemShut {NoStop}%
\bibitem [{\citenamefont {Kurilovich}\ \emph {et~al.}(2021)\citenamefont
  {Kurilovich}, \citenamefont {Kurilovich}, \citenamefont {Fatemi},
  \citenamefont {Devoret},\ and\ \citenamefont {Glazman}}]{Kurilovich2021}%
  \BibitemOpen
  \bibfield  {author} {\bibinfo {author} {\bibfnamefont {P.~D.}\ \bibnamefont
  {Kurilovich}}, \bibinfo {author} {\bibfnamefont {V.~D.}\ \bibnamefont
  {Kurilovich}}, \bibinfo {author} {\bibfnamefont {V.}~\bibnamefont {Fatemi}},
  \bibinfo {author} {\bibfnamefont {M.~H.}\ \bibnamefont {Devoret}},\ and\
  \bibinfo {author} {\bibfnamefont {L.~I.}\ \bibnamefont {Glazman}},\
  }\bibfield  {title} {\bibinfo {title} {Microwave response of an {A}ndreev
  bound state},\ }\href {https://doi.org/10.1103/PhysRevB.104.174517}
  {\bibfield  {journal} {\bibinfo  {journal} {Phys. Rev. B}\ }\textbf {\bibinfo
  {volume} {104}},\ \bibinfo {pages} {174517} (\bibinfo {year}
  {2021})}\BibitemShut {NoStop}%
\bibitem [{\citenamefont {Kurilovich}\ \emph {et~al.}(2024)\citenamefont
  {Kurilovich}, \citenamefont {Kurilovich}, \citenamefont {Svetogorov},
  \citenamefont {Belzig}, \citenamefont {Devoret},\ and\ \citenamefont
  {Glazman}}]{Kurilovich2024}%
  \BibitemOpen
  \bibfield  {author} {\bibinfo {author} {\bibfnamefont {P.~D.}\ \bibnamefont
  {Kurilovich}}, \bibinfo {author} {\bibfnamefont {V.~D.}\ \bibnamefont
  {Kurilovich}}, \bibinfo {author} {\bibfnamefont {A.~E.}\ \bibnamefont
  {Svetogorov}}, \bibinfo {author} {\bibfnamefont {W.}~\bibnamefont {Belzig}},
  \bibinfo {author} {\bibfnamefont {M.~H.}\ \bibnamefont {Devoret}},\ and\
  \bibinfo {author} {\bibfnamefont {L.~I.}\ \bibnamefont {Glazman}},\
  }\bibfield  {title} {\bibinfo {title} {On-demand population of {A}ndreev
  levels by their ionization in the presence of {C}oulomb blockade},\ }\href
  {https://doi.org/10.1103/PhysRevB.110.184508} {\bibfield  {journal} {\bibinfo
   {journal} {Phys. Rev. B}\ }\textbf {\bibinfo {volume} {110}},\ \bibinfo
  {pages} {184508} (\bibinfo {year} {2024})}\BibitemShut {NoStop}%
\bibitem [{\citenamefont {Tosato}\ \emph {et~al.}(2023)\citenamefont {Tosato},
  \citenamefont {Levajac}, \citenamefont {Wang}, \citenamefont {Boor},
  \citenamefont {Borsoi}, \citenamefont {Botifoll}, \citenamefont {Borja},
  \citenamefont {Mart{\'i}-S{\'a}nchez}, \citenamefont {Arbiol}, \citenamefont
  {Sammak}, \citenamefont {Veldhorst},\ and\ \citenamefont
  {Scappucci}}]{Tosato2023}%
  \BibitemOpen
  \bibfield  {author} {\bibinfo {author} {\bibfnamefont {A.}~\bibnamefont
  {Tosato}}, \bibinfo {author} {\bibfnamefont {V.}~\bibnamefont {Levajac}},
  \bibinfo {author} {\bibfnamefont {J.-Y.}\ \bibnamefont {Wang}}, \bibinfo
  {author} {\bibfnamefont {C.~J.}\ \bibnamefont {Boor}}, \bibinfo {author}
  {\bibfnamefont {F.}~\bibnamefont {Borsoi}}, \bibinfo {author} {\bibfnamefont
  {M.}~\bibnamefont {Botifoll}}, \bibinfo {author} {\bibfnamefont {C.~N.}\
  \bibnamefont {Borja}}, \bibinfo {author} {\bibfnamefont {S.}~\bibnamefont
  {Mart{\'i}-S{\'a}nchez}}, \bibinfo {author} {\bibfnamefont {J.}~\bibnamefont
  {Arbiol}}, \bibinfo {author} {\bibfnamefont {A.}~\bibnamefont {Sammak}},
  \bibinfo {author} {\bibfnamefont {M.}~\bibnamefont {Veldhorst}},\ and\
  \bibinfo {author} {\bibfnamefont {G.}~\bibnamefont {Scappucci}},\ }\bibfield
  {title} {\bibinfo {title} {Hard superconducting gap in germanium},\ }\href
  {https://doi.org/10.1038/s43246-023-00351-w} {\bibfield  {journal} {\bibinfo
  {journal} {Commun. Mater.}\ }\textbf {\bibinfo {volume} {4}},\ \bibinfo
  {pages} {23} (\bibinfo {year} {2023})}\BibitemShut {NoStop}%
\bibitem [{\citenamefont {Lodari}\ \emph {et~al.}(2019)\citenamefont {Lodari},
  \citenamefont {Tosato}, \citenamefont {Sabbagh}, \citenamefont {Schubert},
  \citenamefont {Capellini}, \citenamefont {Sammak}, \citenamefont
  {Veldhorst},\ and\ \citenamefont {Scappucci}}]{Lodari2019}%
  \BibitemOpen
  \bibfield  {author} {\bibinfo {author} {\bibfnamefont {M.}~\bibnamefont
  {Lodari}}, \bibinfo {author} {\bibfnamefont {A.}~\bibnamefont {Tosato}},
  \bibinfo {author} {\bibfnamefont {D.}~\bibnamefont {Sabbagh}}, \bibinfo
  {author} {\bibfnamefont {M.~A.}\ \bibnamefont {Schubert}}, \bibinfo {author}
  {\bibfnamefont {G.}~\bibnamefont {Capellini}}, \bibinfo {author}
  {\bibfnamefont {A.}~\bibnamefont {Sammak}}, \bibinfo {author} {\bibfnamefont
  {M.}~\bibnamefont {Veldhorst}},\ and\ \bibinfo {author} {\bibfnamefont
  {G.}~\bibnamefont {Scappucci}},\ }\bibfield  {title} {\bibinfo {title} {Light
  effective hole mass in undoped {G}e/{S}i{G}e quantum wells},\ }\href
  {https://doi.org/10.1103/PhysRevB.100.041304} {\bibfield  {journal} {\bibinfo
   {journal} {Phys. Rev. B}\ }\textbf {\bibinfo {volume} {100}},\ \bibinfo
  {pages} {041304} (\bibinfo {year} {2019})}\BibitemShut {NoStop}%
\bibitem [{\citenamefont {Scappucci}\ \emph {et~al.}(2021)\citenamefont
  {Scappucci}, \citenamefont {Kloeffel}, \citenamefont {Zwanenburg},
  \citenamefont {Loss}, \citenamefont {Myronov}, \citenamefont {Zhang},
  \citenamefont {De~Franceschi}, \citenamefont {Katsaros},\ and\ \citenamefont
  {Veldhorst}}]{Scappucci2021}%
  \BibitemOpen
  \bibfield  {author} {\bibinfo {author} {\bibfnamefont {G.}~\bibnamefont
  {Scappucci}}, \bibinfo {author} {\bibfnamefont {C.}~\bibnamefont {Kloeffel}},
  \bibinfo {author} {\bibfnamefont {F.~A.}\ \bibnamefont {Zwanenburg}},
  \bibinfo {author} {\bibfnamefont {D.}~\bibnamefont {Loss}}, \bibinfo {author}
  {\bibfnamefont {M.}~\bibnamefont {Myronov}}, \bibinfo {author} {\bibfnamefont
  {J.-J.}\ \bibnamefont {Zhang}}, \bibinfo {author} {\bibfnamefont
  {S.}~\bibnamefont {De~Franceschi}}, \bibinfo {author} {\bibfnamefont
  {G.}~\bibnamefont {Katsaros}},\ and\ \bibinfo {author} {\bibfnamefont
  {M.}~\bibnamefont {Veldhorst}},\ }\bibfield  {title} {\bibinfo {title} {The
  germanium quantum information route},\ }\href
  {https://doi.org/10.1038/s41578-020-00262-z} {\bibfield  {journal} {\bibinfo
  {journal} {Nat. Rev. Mater.}\ }\textbf {\bibinfo {volume} {6}},\ \bibinfo
  {pages} {926} (\bibinfo {year} {2021})}\BibitemShut {NoStop}%
\bibitem [{\citenamefont {Lakic}\ \emph {et~al.}(2025)\citenamefont {Lakic},
  \citenamefont {Lawrie}, \citenamefont {van Driel}, \citenamefont {Stehouwer},
  \citenamefont {Su}, \citenamefont {Veldhorst}, \citenamefont {Scappucci},
  \citenamefont {Kuemmeth},\ and\ \citenamefont {Chatterjee}}]{Lakic2025}%
  \BibitemOpen
  \bibfield  {author} {\bibinfo {author} {\bibfnamefont {L.}~\bibnamefont
  {Lakic}}, \bibinfo {author} {\bibfnamefont {W.~I.~L.}\ \bibnamefont
  {Lawrie}}, \bibinfo {author} {\bibfnamefont {D.}~\bibnamefont {van Driel}},
  \bibinfo {author} {\bibfnamefont {L.~E.~A.}\ \bibnamefont {Stehouwer}},
  \bibinfo {author} {\bibfnamefont {Y.}~\bibnamefont {Su}}, \bibinfo {author}
  {\bibfnamefont {M.}~\bibnamefont {Veldhorst}}, \bibinfo {author}
  {\bibfnamefont {G.}~\bibnamefont {Scappucci}}, \bibinfo {author}
  {\bibfnamefont {F.}~\bibnamefont {Kuemmeth}},\ and\ \bibinfo {author}
  {\bibfnamefont {A.}~\bibnamefont {Chatterjee}},\ }\bibfield  {title}
  {\bibinfo {title} {A quantum dot in germanium proximitized by a
  superconductor},\ }\href {https://doi.org/10.1038/s41563-024-02095-5}
  {\bibfield  {journal} {\bibinfo  {journal} {Nat. Mater.}\ }\textbf {\bibinfo
  {volume} {24}},\ \bibinfo {pages} {552} (\bibinfo {year} {2025})}\BibitemShut
  {NoStop}%
\bibitem [{\citenamefont {Hinderling}\ \emph
  {et~al.}(2024{\natexlab{b}})\citenamefont {Hinderling}, \citenamefont {Kate},
  \citenamefont {Haxell}, \citenamefont {Coraiola}, \citenamefont {Paredes},
  \citenamefont {Cheah}, \citenamefont {Krizek}, \citenamefont {Schott},
  \citenamefont {Wegscheider}, \citenamefont {Sabonis},\ and\ \citenamefont
  {Nichele}}]{Hinderling2024}%
  \BibitemOpen
  \bibfield  {author} {\bibinfo {author} {\bibfnamefont {M.}~\bibnamefont
  {Hinderling}}, \bibinfo {author} {\bibfnamefont {S.~t.}\ \bibnamefont
  {Kate}}, \bibinfo {author} {\bibfnamefont {D.}~\bibnamefont {Haxell}},
  \bibinfo {author} {\bibfnamefont {M.}~\bibnamefont {Coraiola}}, \bibinfo
  {author} {\bibfnamefont {S.}~\bibnamefont {Paredes}}, \bibinfo {author}
  {\bibfnamefont {E.}~\bibnamefont {Cheah}}, \bibinfo {author} {\bibfnamefont
  {F.}~\bibnamefont {Krizek}}, \bibinfo {author} {\bibfnamefont
  {R.}~\bibnamefont {Schott}}, \bibinfo {author} {\bibfnamefont
  {W.}~\bibnamefont {Wegscheider}}, \bibinfo {author} {\bibfnamefont
  {D.}~\bibnamefont {Sabonis}},\ and\ \bibinfo {author} {\bibfnamefont
  {F.}~\bibnamefont {Nichele}},\ }\bibfield  {title} {\bibinfo {title}
  {Flip-chip-based fast inductive parity readout of a planar superconducting
  island},\ }\href {https://doi.org/10.1103/PRXQuantum.5.030337} {\bibfield
  {journal} {\bibinfo  {journal} {PRX Quantum}\ }\textbf {\bibinfo {volume}
  {5}},\ \bibinfo {pages} {030337} (\bibinfo {year}
  {2024}{\natexlab{b}})}\BibitemShut {NoStop}%
\bibitem [{\citenamefont {Probst}\ \emph {et~al.}(2015)\citenamefont {Probst},
  \citenamefont {Song}, \citenamefont {Bushev}, \citenamefont {Ustinov},\ and\
  \citenamefont {Weides}}]{Probst2015}%
  \BibitemOpen
  \bibfield  {author} {\bibinfo {author} {\bibfnamefont {S.}~\bibnamefont
  {Probst}}, \bibinfo {author} {\bibfnamefont {F.~B.}\ \bibnamefont {Song}},
  \bibinfo {author} {\bibfnamefont {P.~A.}\ \bibnamefont {Bushev}}, \bibinfo
  {author} {\bibfnamefont {A.~V.}\ \bibnamefont {Ustinov}},\ and\ \bibinfo
  {author} {\bibfnamefont {M.}~\bibnamefont {Weides}},\ }\bibfield  {title}
  {\bibinfo {title} {Efficient and robust analysis of complex scattering data
  under noise in microwave resonators},\ }\href
  {https://doi.org/10.1063/1.4907935} {\bibfield  {journal} {\bibinfo
  {journal} {Rev. Sci. Instrum.}\ }\textbf {\bibinfo {volume} {86}},\ \bibinfo
  {pages} {024706} (\bibinfo {year} {2015})}\BibitemShut {NoStop}%
\bibitem [{\citenamefont {Bedell}\ \emph {et~al.}(2020)\citenamefont {Bedell},
  \citenamefont {Hart}, \citenamefont {Bangsaruntip}, \citenamefont {Durfee},
  \citenamefont {Ott}, \citenamefont {Hopstaken}, \citenamefont {Carroll},\
  and\ \citenamefont {Gumann}}]{Bedell2020}%
  \BibitemOpen
  \bibfield  {author} {\bibinfo {author} {\bibfnamefont {S.~W.}\ \bibnamefont
  {Bedell}}, \bibinfo {author} {\bibfnamefont {S.}~\bibnamefont {Hart}},
  \bibinfo {author} {\bibfnamefont {S.}~\bibnamefont {Bangsaruntip}}, \bibinfo
  {author} {\bibfnamefont {C.}~\bibnamefont {Durfee}}, \bibinfo {author}
  {\bibfnamefont {J.~A.}\ \bibnamefont {Ott}}, \bibinfo {author} {\bibfnamefont
  {M.}~\bibnamefont {Hopstaken}}, \bibinfo {author} {\bibfnamefont {M.~S.}\
  \bibnamefont {Carroll}},\ and\ \bibinfo {author} {\bibfnamefont
  {P.}~\bibnamefont {Gumann}},\ }\bibfield  {title} {\bibinfo {title}
  {(invited) low-temperature growth of strained germanium quantum wells for
  high mobility applications},\ }\href {https://doi.org/10.1149/09805.0215ecst}
  {\bibfield  {journal} {\bibinfo  {journal} {ECS Trans.}\ }\textbf {\bibinfo
  {volume} {98}},\ \bibinfo {pages} {215} (\bibinfo {year} {2020})}\BibitemShut
  {NoStop}%
\bibitem [{\citenamefont {Watanabe}\ \emph {et~al.}(2009)\citenamefont
  {Watanabe}, \citenamefont {Inomata}, \citenamefont {Yamamoto},\ and\
  \citenamefont {Tsai}}]{Watanabe2009}%
  \BibitemOpen
  \bibfield  {author} {\bibinfo {author} {\bibfnamefont {M.}~\bibnamefont
  {Watanabe}}, \bibinfo {author} {\bibfnamefont {K.}~\bibnamefont {Inomata}},
  \bibinfo {author} {\bibfnamefont {T.}~\bibnamefont {Yamamoto}},\ and\
  \bibinfo {author} {\bibfnamefont {J.-S.}\ \bibnamefont {Tsai}},\ }\bibfield
  {title} {\bibinfo {title} {Power-dependent internal loss in {J}osephson
  bifurcation amplifiers},\ }\href {https://doi.org/10.1103/PhysRevB.80.174502}
  {\bibfield  {journal} {\bibinfo  {journal} {Phys. Rev. B}\ }\textbf {\bibinfo
  {volume} {80}},\ \bibinfo {pages} {174502} (\bibinfo {year}
  {2009})}\BibitemShut {NoStop}%
\bibitem [{\citenamefont {Romero}\ \emph {et~al.}(2012)\citenamefont {Romero},
  \citenamefont {Lizuain}, \citenamefont {Shumeiko}, \citenamefont {Solano},\
  and\ \citenamefont {Bergeret}}]{Romero2012}%
  \BibitemOpen
  \bibfield  {author} {\bibinfo {author} {\bibfnamefont {G.}~\bibnamefont
  {Romero}}, \bibinfo {author} {\bibfnamefont {I.}~\bibnamefont {Lizuain}},
  \bibinfo {author} {\bibfnamefont {V.~S.}\ \bibnamefont {Shumeiko}}, \bibinfo
  {author} {\bibfnamefont {E.}~\bibnamefont {Solano}},\ and\ \bibinfo {author}
  {\bibfnamefont {F.~S.}\ \bibnamefont {Bergeret}},\ }\bibfield  {title}
  {\bibinfo {title} {Circuit quantum electrodynamics with a superconducting
  quantum point contact},\ }\href {https://doi.org/10.1103/PhysRevB.85.180506}
  {\bibfield  {journal} {\bibinfo  {journal} {Phys. Rev. B}\ }\textbf {\bibinfo
  {volume} {85}},\ \bibinfo {pages} {180506} (\bibinfo {year}
  {2012})}\BibitemShut {NoStop}%
\bibitem [{\citenamefont {Park}\ \emph {et~al.}(2020)\citenamefont {Park},
  \citenamefont {Metzger}, \citenamefont {Tosi}, \citenamefont {Goffman},
  \citenamefont {Urbina}, \citenamefont {Pothier},\ and\ \citenamefont
  {Yeyati}}]{Park2020}%
  \BibitemOpen
  \bibfield  {author} {\bibinfo {author} {\bibfnamefont {S.}~\bibnamefont
  {Park}}, \bibinfo {author} {\bibfnamefont {C.}~\bibnamefont {Metzger}},
  \bibinfo {author} {\bibfnamefont {L.}~\bibnamefont {Tosi}}, \bibinfo {author}
  {\bibfnamefont {M.~F.}\ \bibnamefont {Goffman}}, \bibinfo {author}
  {\bibfnamefont {C.}~\bibnamefont {Urbina}}, \bibinfo {author} {\bibfnamefont
  {H.}~\bibnamefont {Pothier}},\ and\ \bibinfo {author} {\bibfnamefont {A.~L.}\
  \bibnamefont {Yeyati}},\ }\bibfield  {title} {\bibinfo {title} {From
  adiabatic to dispersive readout of quantum circuits},\ }\href
  {https://doi.org/10.1103/PhysRevLett.125.077701} {\bibfield  {journal}
  {\bibinfo  {journal} {Phys. Rev. Lett.}\ }\textbf {\bibinfo {volume} {125}},\
  \bibinfo {pages} {077701} (\bibinfo {year} {2020})}\BibitemShut {NoStop}%
\bibitem [{\citenamefont {Metzger}\ \emph {et~al.}(2021)\citenamefont
  {Metzger}, \citenamefont {Park}, \citenamefont {Tosi}, \citenamefont
  {Janvier}, \citenamefont {Reynoso}, \citenamefont {Goffman}, \citenamefont
  {Urbina}, \citenamefont {Levy~Yeyati},\ and\ \citenamefont
  {Pothier}}]{Metzger2021}%
  \BibitemOpen
  \bibfield  {author} {\bibinfo {author} {\bibfnamefont {C.}~\bibnamefont
  {Metzger}}, \bibinfo {author} {\bibfnamefont {S.}~\bibnamefont {Park}},
  \bibinfo {author} {\bibfnamefont {L.}~\bibnamefont {Tosi}}, \bibinfo {author}
  {\bibfnamefont {C.}~\bibnamefont {Janvier}}, \bibinfo {author} {\bibfnamefont
  {A.~A.}\ \bibnamefont {Reynoso}}, \bibinfo {author} {\bibfnamefont {M.~F.}\
  \bibnamefont {Goffman}}, \bibinfo {author} {\bibfnamefont {C.}~\bibnamefont
  {Urbina}}, \bibinfo {author} {\bibfnamefont {A.}~\bibnamefont
  {Levy~Yeyati}},\ and\ \bibinfo {author} {\bibfnamefont {H.}~\bibnamefont
  {Pothier}},\ }\bibfield  {title} {\bibinfo {title} {Circuit-{Q}{E}{D} with
  phase-biased {J}osephson weak links},\ }\href
  {https://doi.org/10.1103/PhysRevResearch.3.013036} {\bibfield  {journal}
  {\bibinfo  {journal} {Phys. Rev. Res.}\ }\textbf {\bibinfo {volume} {3}},\
  \bibinfo {pages} {013036} (\bibinfo {year} {2021})}\BibitemShut {NoStop}%
\bibitem [{\citenamefont {Haller}\ \emph {et~al.}(2022)\citenamefont {Haller},
  \citenamefont {F\"ul\"op}, \citenamefont {Indolese}, \citenamefont
  {Ridderbos}, \citenamefont {Kraft}, \citenamefont {Cheung}, \citenamefont
  {Ungerer}, \citenamefont {Watanabe}, \citenamefont {Taniguchi}, \citenamefont
  {Beckmann}, \citenamefont {Danneau}, \citenamefont {Virtanen},\ and\
  \citenamefont {Sch\"onenberger}}]{Haller2022}%
  \BibitemOpen
  \bibfield  {author} {\bibinfo {author} {\bibfnamefont {R.}~\bibnamefont
  {Haller}}, \bibinfo {author} {\bibfnamefont {G.}~\bibnamefont {F\"ul\"op}},
  \bibinfo {author} {\bibfnamefont {D.}~\bibnamefont {Indolese}}, \bibinfo
  {author} {\bibfnamefont {J.}~\bibnamefont {Ridderbos}}, \bibinfo {author}
  {\bibfnamefont {R.}~\bibnamefont {Kraft}}, \bibinfo {author} {\bibfnamefont
  {L.~Y.}\ \bibnamefont {Cheung}}, \bibinfo {author} {\bibfnamefont {J.~H.}\
  \bibnamefont {Ungerer}}, \bibinfo {author} {\bibfnamefont {K.}~\bibnamefont
  {Watanabe}}, \bibinfo {author} {\bibfnamefont {T.}~\bibnamefont {Taniguchi}},
  \bibinfo {author} {\bibfnamefont {D.}~\bibnamefont {Beckmann}}, \bibinfo
  {author} {\bibfnamefont {R.}~\bibnamefont {Danneau}}, \bibinfo {author}
  {\bibfnamefont {P.}~\bibnamefont {Virtanen}},\ and\ \bibinfo {author}
  {\bibfnamefont {C.}~\bibnamefont {Sch\"onenberger}},\ }\bibfield  {title}
  {\bibinfo {title} {Phase-dependent microwave response of a graphene
  {J}osephson junction},\ }\href
  {https://doi.org/10.1103/PhysRevResearch.4.013198} {\bibfield  {journal}
  {\bibinfo  {journal} {Phys. Rev. Res.}\ }\textbf {\bibinfo {volume} {4}},\
  \bibinfo {pages} {013198} (\bibinfo {year} {2022})}\BibitemShut {NoStop}%
\bibitem [{\citenamefont {Bagwell}(1992)}]{Bagwell1992}%
  \BibitemOpen
  \bibfield  {author} {\bibinfo {author} {\bibfnamefont {P.~F.}\ \bibnamefont
  {Bagwell}},\ }\bibfield  {title} {\bibinfo {title} {Suppression of the
  {J}osephson current through a narrow, mesoscopic, semiconductor channel by a
  single impurity},\ }\href {https://doi.org/10.1103/PhysRevB.46.12573}
  {\bibfield  {journal} {\bibinfo  {journal} {Phys. Rev. B}\ }\textbf {\bibinfo
  {volume} {46}},\ \bibinfo {pages} {12573} (\bibinfo {year}
  {1992})}\BibitemShut {NoStop}%
\bibitem [{\citenamefont {Sahu}\ \emph {et~al.}(2024)\citenamefont {Sahu},
  \citenamefont {Matute-Ca\~nadas}, \citenamefont {Benito}, \citenamefont
  {Krogstrup}, \citenamefont {Nyg\aa{}rd}, \citenamefont {Goffman},
  \citenamefont {Urbina}, \citenamefont {Yeyati},\ and\ \citenamefont
  {Pothier}}]{Sahu2024}%
  \BibitemOpen
  \bibfield  {author} {\bibinfo {author} {\bibfnamefont {M.~R.}\ \bibnamefont
  {Sahu}}, \bibinfo {author} {\bibfnamefont {F.~J.}\ \bibnamefont
  {Matute-Ca\~nadas}}, \bibinfo {author} {\bibfnamefont {M.}~\bibnamefont
  {Benito}}, \bibinfo {author} {\bibfnamefont {P.}~\bibnamefont {Krogstrup}},
  \bibinfo {author} {\bibfnamefont {J.}~\bibnamefont {Nyg\aa{}rd}}, \bibinfo
  {author} {\bibfnamefont {M.~F.}\ \bibnamefont {Goffman}}, \bibinfo {author}
  {\bibfnamefont {C.}~\bibnamefont {Urbina}}, \bibinfo {author} {\bibfnamefont
  {A.~L.}\ \bibnamefont {Yeyati}},\ and\ \bibinfo {author} {\bibfnamefont
  {H.}~\bibnamefont {Pothier}},\ }\bibfield  {title} {\bibinfo {title}
  {Ground-state phase diagram and parity-flipping microwave transitions in a
  gate-tunable {J}osephson junction},\ }\href
  {https://doi.org/10.1103/PhysRevB.109.134506} {\bibfield  {journal} {\bibinfo
   {journal} {Phys. Rev. B}\ }\textbf {\bibinfo {volume} {109}},\ \bibinfo
  {pages} {134506} (\bibinfo {year} {2024})}\BibitemShut {NoStop}%
\bibitem [{\citenamefont {Massai}\ \emph {et~al.}(2024)\citenamefont {Massai},
  \citenamefont {Het{\'e}nyi}, \citenamefont {Mergenthaler}, \citenamefont
  {Schupp}, \citenamefont {Sommer}, \citenamefont {Paredes}, \citenamefont
  {Bedell}, \citenamefont {Harvey-Collard}, \citenamefont {Salis},
  \citenamefont {Fuhrer},\ and\ \citenamefont {Hendrickx}}]{Massai2024}%
  \BibitemOpen
  \bibfield  {author} {\bibinfo {author} {\bibfnamefont {L.}~\bibnamefont
  {Massai}}, \bibinfo {author} {\bibfnamefont {B.}~\bibnamefont {Het{\'e}nyi}},
  \bibinfo {author} {\bibfnamefont {M.}~\bibnamefont {Mergenthaler}}, \bibinfo
  {author} {\bibfnamefont {F.~J.}\ \bibnamefont {Schupp}}, \bibinfo {author}
  {\bibfnamefont {L.}~\bibnamefont {Sommer}}, \bibinfo {author} {\bibfnamefont
  {S.}~\bibnamefont {Paredes}}, \bibinfo {author} {\bibfnamefont {S.~W.}\
  \bibnamefont {Bedell}}, \bibinfo {author} {\bibfnamefont {P.}~\bibnamefont
  {Harvey-Collard}}, \bibinfo {author} {\bibfnamefont {G.}~\bibnamefont
  {Salis}}, \bibinfo {author} {\bibfnamefont {A.}~\bibnamefont {Fuhrer}},\ and\
  \bibinfo {author} {\bibfnamefont {N.~W.}\ \bibnamefont {Hendrickx}},\
  }\bibfield  {title} {\bibinfo {title} {Impact of interface traps on charge
  noise and low-density transport properties in {G}e/{S}i{G}e
  heterostructures},\ }\href {https://doi.org/10.1038/s43246-024-00563-8}
  {\bibfield  {journal} {\bibinfo  {journal} {Commun. Mater.}\ }\textbf
  {\bibinfo {volume} {5}},\ \bibinfo {pages} {151} (\bibinfo {year}
  {2024})}\BibitemShut {NoStop}%
\bibitem [{\citenamefont {Bosco}\ \emph {et~al.}(2021)\citenamefont {Bosco},
  \citenamefont {Benito}, \citenamefont {Adelsberger},\ and\ \citenamefont
  {Loss}}]{Bosco2021}%
  \BibitemOpen
  \bibfield  {author} {\bibinfo {author} {\bibfnamefont {S.}~\bibnamefont
  {Bosco}}, \bibinfo {author} {\bibfnamefont {M.}~\bibnamefont {Benito}},
  \bibinfo {author} {\bibfnamefont {C.}~\bibnamefont {Adelsberger}},\ and\
  \bibinfo {author} {\bibfnamefont {D.}~\bibnamefont {Loss}},\ }\bibfield
  {title} {\bibinfo {title} {Squeezed hole spin qubits in {G}e quantum dots
  with ultrafast gates at low power},\ }\href
  {https://doi.org/10.1103/PhysRevB.104.115425} {\bibfield  {journal} {\bibinfo
   {journal} {Phys. Rev. B}\ }\textbf {\bibinfo {volume} {104}},\ \bibinfo
  {pages} {115425} (\bibinfo {year} {2021})}\BibitemShut {NoStop}%
\bibitem [{\citenamefont {Adelsberger}\ \emph {et~al.}(2022)\citenamefont
  {Adelsberger}, \citenamefont {Bosco}, \citenamefont {Klinovaja},\ and\
  \citenamefont {Loss}}]{Adelsberger2022}%
  \BibitemOpen
  \bibfield  {author} {\bibinfo {author} {\bibfnamefont {C.}~\bibnamefont
  {Adelsberger}}, \bibinfo {author} {\bibfnamefont {S.}~\bibnamefont {Bosco}},
  \bibinfo {author} {\bibfnamefont {J.}~\bibnamefont {Klinovaja}},\ and\
  \bibinfo {author} {\bibfnamefont {D.}~\bibnamefont {Loss}},\ }\bibfield
  {title} {\bibinfo {title} {Enhanced orbital magnetic field effects in {G}e
  hole nanowires},\ }\href {https://doi.org/10.1103/PhysRevB.106.235408}
  {\bibfield  {journal} {\bibinfo  {journal} {Phys. Rev. B}\ }\textbf {\bibinfo
  {volume} {106}},\ \bibinfo {pages} {235408} (\bibinfo {year}
  {2022})}\BibitemShut {NoStop}%
\bibitem [{\citenamefont {Hoffman}\ and\ \citenamefont
  {Tahan}(2025)}]{Hoffman2025}%
  \BibitemOpen
  \bibfield  {author} {\bibinfo {author} {\bibfnamefont {S.}~\bibnamefont
  {Hoffman}}\ and\ \bibinfo {author} {\bibfnamefont {C.}~\bibnamefont
  {Tahan}},\ }\href {https://doi.org/10.48550/arXiv.2506.13988} {\bibinfo
  {title} {Resolving {A}ndreev spin qubits in germanium-based {J}osephson
  junctions}},\ \bibinfo {howpublished} {arXiv:2506.13988} (\bibinfo {year}
  {2025})\BibitemShut {NoStop}%
\bibitem [{\citenamefont {Luethi}\ \emph {et~al.}(2023)\citenamefont {Luethi},
  \citenamefont {Laubscher}, \citenamefont {Bosco}, \citenamefont {Loss},\ and\
  \citenamefont {Klinovaja}}]{Luethi2023}%
  \BibitemOpen
  \bibfield  {author} {\bibinfo {author} {\bibfnamefont {M.}~\bibnamefont
  {Luethi}}, \bibinfo {author} {\bibfnamefont {K.}~\bibnamefont {Laubscher}},
  \bibinfo {author} {\bibfnamefont {S.}~\bibnamefont {Bosco}}, \bibinfo
  {author} {\bibfnamefont {D.}~\bibnamefont {Loss}},\ and\ \bibinfo {author}
  {\bibfnamefont {J.}~\bibnamefont {Klinovaja}},\ }\bibfield  {title} {\bibinfo
  {title} {Planar {J}osephson junctions in germanium: {E}ffect of cubic
  spin-orbit interaction},\ }\href
  {https://doi.org/10.1103/PhysRevB.107.035435} {\bibfield  {journal} {\bibinfo
   {journal} {Phys. Rev. B}\ }\textbf {\bibinfo {volume} {107}},\ \bibinfo
  {pages} {035435} (\bibinfo {year} {2023})}\BibitemShut {NoStop}%
\bibitem [{\citenamefont {Costa}\ \emph {et~al.}(2025)\citenamefont {Costa},
  \citenamefont {Vecchio}, \citenamefont {Hudson}, \citenamefont {Stehouwer},
  \citenamefont {Tosato}, \citenamefont {Esposti}, \citenamefont {Lodari},
  \citenamefont {Bosco},\ and\ \citenamefont {Scappucci}}]{Costa2025}%
  \BibitemOpen
  \bibfield  {author} {\bibinfo {author} {\bibfnamefont {D.}~\bibnamefont
  {Costa}}, \bibinfo {author} {\bibfnamefont {P.~D.}\ \bibnamefont {Vecchio}},
  \bibinfo {author} {\bibfnamefont {K.}~\bibnamefont {Hudson}}, \bibinfo
  {author} {\bibfnamefont {L.~E.~A.}\ \bibnamefont {Stehouwer}}, \bibinfo
  {author} {\bibfnamefont {A.}~\bibnamefont {Tosato}}, \bibinfo {author}
  {\bibfnamefont {D.~D.}\ \bibnamefont {Esposti}}, \bibinfo {author}
  {\bibfnamefont {M.}~\bibnamefont {Lodari}}, \bibinfo {author} {\bibfnamefont
  {S.}~\bibnamefont {Bosco}},\ and\ \bibinfo {author} {\bibfnamefont
  {G.}~\bibnamefont {Scappucci}},\ }\href
  {https://doi.org/10.48550/arXiv.2506.04724} {\bibinfo {title} {Buried
  unstrained germanium channels: a lattice-matched platform for quantum
  technology}},\ \bibinfo {howpublished} {arXiv:2506.04724} (\bibinfo {year}
  {2025})\BibitemShut {NoStop}%
\bibitem [{\citenamefont {Shvetsov}\ \emph {et~al.}(2025)\citenamefont
  {Shvetsov}, \citenamefont {Khola}, \citenamefont {Buccheri}, \citenamefont
  {Cools}, \citenamefont {Trnjanin}, \citenamefont {Kanne}, \citenamefont
  {Nygård},\ and\ \citenamefont {Geresdi}}]{Shvetsov2025}%
  \BibitemOpen
  \bibfield  {author} {\bibinfo {author} {\bibfnamefont {O.~O.}\ \bibnamefont
  {Shvetsov}}, \bibinfo {author} {\bibfnamefont {A.}~\bibnamefont {Khola}},
  \bibinfo {author} {\bibfnamefont {V.}~\bibnamefont {Buccheri}}, \bibinfo
  {author} {\bibfnamefont {I.~P.~C.}\ \bibnamefont {Cools}}, \bibinfo {author}
  {\bibfnamefont {N.}~\bibnamefont {Trnjanin}}, \bibinfo {author}
  {\bibfnamefont {T.}~\bibnamefont {Kanne}}, \bibinfo {author} {\bibfnamefont
  {J.}~\bibnamefont {Nygård}},\ and\ \bibinfo {author} {\bibfnamefont
  {A.}~\bibnamefont {Geresdi}},\ }\href
  {https://doi.org/10.48550/arXiv.2502.09243} {\bibinfo {title} {Approaching
  the ultrastrong coupling regime between an {A}ndreev level and a microwave
  resonator}},\ \bibinfo {howpublished} {arXiv:2502.09243} (\bibinfo {year}
  {2025})\BibitemShut {NoStop}%
\bibitem [{\citenamefont {Park}\ and\ \citenamefont
  {Yeyati}(2017)}]{Yeyati2017}%
  \BibitemOpen
  \bibfield  {author} {\bibinfo {author} {\bibfnamefont {S.}~\bibnamefont
  {Park}}\ and\ \bibinfo {author} {\bibfnamefont {A.~L.}\ \bibnamefont
  {Yeyati}},\ }\bibfield  {title} {\bibinfo {title} {{A}ndreev spin qubits in
  multichannel {R}ashba nanowires},\ }\href
  {https://doi.org/10.1103/PhysRevB.96.125416} {\bibfield  {journal} {\bibinfo
  {journal} {Phys. Rev. B}\ }\textbf {\bibinfo {volume} {96}},\ \bibinfo
  {pages} {125416} (\bibinfo {year} {2017})}\BibitemShut {NoStop}%
\end{thebibliography}%

\newpage
\setcounter{section}{0}
\onecolumngrid

\newcounter{myc}
\renewcommand{\thefigure}{S.\arabic{myc}}

\newcounter{mye}
\renewcommand{\theequation}{S.\arabic{mye}}

\newpage	
\setlength{\parskip}{0pt}

\end{document}